\documentclass[fleqn,usenatbib]{mnras}

\usepackage[T1]{fontenc}
\usepackage{ae,aecompl}
\usepackage{multirow}
\usepackage{tabularx}
\usepackage{supertabular}
\usepackage{longtable}
\usepackage[]{url}
\usepackage{graphicx}	
\usepackage{amsmath}	
\usepackage{amssymb}	

\usepackage{graphicx}	
\usepackage{amsmath}	
\usepackage{amssymb}	
\usepackage{multirow}
\usepackage[left]{lineno}
\usepackage{booktabs}
\usepackage{gensymb}
\usepackage{xcolor,color}
\usepackage[caption = false]{subfig}
\usepackage{ulem}

\newcommand{\ztfink}[1]{ZTF-Fink}
\newcommand{\fink}{{\sc Fink}}

\newcommand{\lt}{\ensuremath <}

\title{GRANDMA Observations of ZTF/\textit{Fink} Transients during Summer 2021}

\author[GRANDMA consortium]{
V. Aivazyan$^{1,2}$,
M. Almualla$^{3}$,
S. Antier$^{4,5,6}$
A. Baransky$^{7}$, 
K. Barynova$^{8}$, 
\newauthor
S. Basa$^{9}$,
F. Bayard$^{10}$,
S. Beradze$^{1,2}$,
D. Berezin$^{11}$, 
M. Blazek$^{12}$, 
\newauthor
D. Boutigny$^{13}$,
D. Boust$^{14,15}$,
E. Broens$^{16,17}$,
O. Burkhonov$^{18}$,
A. Cailleau$^{19}$,
\newauthor
N. Christensen$^{4}$,
D. Cejudo$^{20}$,
A. Coleiro$^{6}$,
M. W. Coughlin$^{21}$,
D. Datashvili$^{1,2}$,
\newauthor
T. Dietrich$^{22,23}$, 
F. Dolon$^{24}$,
J.-G. Ducoin$^{25,26}$,
P.-A. Duverne$^{25}$,
G. Marchal-Duval$^{25}$,
\newauthor
C. Galdies$^{27,28}$,
L. Granier$^{29}$,
V. Godunova$^{11}$, 
P. Gokuldass$^{30}$ 
H. B. Eggenstein$^{31}$,
\newauthor
M. Freeberg$^{32}$,
P. Hello$^{25}$,
R. Inasaridze$^{1,2}$,
E. E. O. Ishida$^{33}$,
P. Jaquiery$^{34}$,
\newauthor
D. A. Kann$^{12}$,
G. Kapanadze$^{1,2}$,
S. Karpov$^{35}$,
R. W. Kiendrebeogo$^{4,36}$
A. Klotz$^{37,38}$,
\newauthor
R. Kneip$^{39}$,
N. Kochiashvili$^{1}$,
W. Kou$^{40}$,
F. Kugel$^{14}$,
C. Lachaud$^{6}$,
S. Leonini$^{41}$,
\newauthor
A. Leroy$^{42,43}$,
N. Leroy$^{25}$,
A. Le Van Su$^{24}$,
D. Marchais$^{44}$,
M. Ma\v{s}ek$^{35}$,
T. Midavaine$^{6}$,
\newauthor
A. Möller$^{33,45}$,
D. Morris$^{30}$,
R. Natsvlishvili$^{1}$,
F. Navarete$^{46}$,
K. Noysena$^{47}$,
S. Nissanke$^{5}$,
\newauthor
K. Noonan$^{48}$
N. B. Orange$^{48}$,
J. Peloton$^{25}$,
A. Popowicz$^{49}$,
T. Pradier$^{50}$,
\newauthor
M. Prouza$^{35}$,
G. Raaijmakers$^{5}$,
Y. Rajabov$^{18}$,
M. Richmond$^{51}$,
Ya. Romanyuk$^{52}$, 
\newauthor
L. Rousselot$^{53}$,
T. Sadibekova$^{18, 53}$,
M. Serrau$^{14}$,
O. Sokoliuk$^{7,52}$,
X. Song$^{40}$,
\newauthor
A. Simon $^{55,56}$, 
C. Stachie$^{4}$,
A. Taylor $^{57,58,59}$,
Y. Tillayev$^{18,60}$,
D. Turpin$^{54}$\footnote{damien.turpin@cea.fr},
\newauthor
M. Vardosanidze$^{1,2}$,
J. Vlieghe$^{25}$,
I. Tosta e Melo$^{61}$,
X. F. Wang$^{62,40}$,
J. Zhu$^{40}$
}
\date{Accepted XXX. Received YYY; in original form ZZZ}
\pubyear{2022}

\begin{document}
\label{firstpage}
\pagerange{\pageref{firstpage}--\pageref{lastpage}}
\maketitle

\begin{abstract}
We present our follow-up observations with GRANDMA of transient sources revealed by the Zwicky Transient Facility (ZTF). Over a period of six months, from 1 April to 30 September 2021, all ZTF triggers were examined in real time by a dedicated science module implemented in the \textit{Fink} broker, which will be used for the data processing of the Vera C. Rubin Observatory. In this article, we present three selection methods to identify promising kilonova candidates. Out of more than 35 million candidates, a hundred sources have passed our selection criteria. Six were then followed-up by GRANDMA (by both professional and amateur astronomers). The majority were finally classified either as asteroids or as supernovae events. We then demonstrate the added value of a substantial follow-up campaign of optical transients. We mobilized 37 telescopes, bringing together a large sample of images, taken under various conditions and quality. To complement the orphan kilonova candidates (those without associated gamma-ray bursts, which were all), we included three additional supernovae alerts to conduct further observations of during summer 2021. We demonstrate the importance of the amateur astronomer community that contributed from one to 23 images within the first two days after the \textit{Fink} alert for scientific analyzes of new sources discovered in a magnitude range $r^\prime=17-19$ mag. We based our rapid kilonova classification on the decay rate of the optical source that should exceed 0.3 mag/day. GRANDMA's follow-up determined the fading rate within $1.5\pm1.2$ days post-discovery, without waiting for further observations from ZTF. No confirmed kilonovae were discovered during our observing campaign. This work will be continued in the coming months in the view of preparing for kilonova searches in the next gravitational-wave observing run O4 and the commissioning of the Vera C. Rubin Observatory. 
\end{abstract}




\section{Introduction}

The landmark detection of GW170817 by LIGO and Virgo \citep{LSC_BNS_2017PhRvL}, GRB 170817A by Fermi-GBM~\citep{goldstein_ordinary_2017} and INTEGRAL~\citep{savchenko_integral_2017}, and the subsequent accompanying host of electromagnetic signatures~\citep{LSC_MM_2017ApJ} solidified the long predicted connection of binary neutron star mergers (BNS) to short gamma-ray bursts (sGRBs; e.g., \citealt{2017ApJ...848L..21A, 2017ApJ...848L..25H, Hallinan_2017Sci}), and optical/infrared transients called kilonovae (KNe; e.g., \citealt{Andreoni_2017PASA, 2017Natur.551..210A, 2017SciBu..62.1433H}). Signatures of KNe also exist in some sGRBs (e.g., \citealt{2013Natur.500..547T,2013ApJ...774L..23B}), and efforts to obtain a full characterization of their emission remain a major priority in Astrophysics. In particular, KN emission is uniquely suited to better constrain the neutron star equation of state (EoS, e.g., \citealt{2017ApJ...850L..34B, 2017ApJ...850L..19M,CoDi2018,CoDi2018b,Coughlin2019}), the Hubble constant (e.g., \citealt{LSC_BNS_2017PhRvL,2019NatAs...3..940H,2020Sci...370.1450D, PhysRevResearch.2.022006,2020NatCo..11.4129C}) and the abundancies of $r$-process nucleosynthesis (e.g., \citealt{2017ApJ...848L..19C, 2017Sci...358.1556C,2017ApJ...848L..17C}).

To address outstanding questions related to KNe, e.g., answering whether or not diverse processes take place in their ejecta and the sources of different emission components, a large sample of sources are necessary \citep{2021arXiv210606820A,2021ApJ...918...63A}. However, obtaining such a large data set is a difficult task for several reasons. Firstly, KNe are rare ($\lt$\,1\% of the core collapse supernova rate), fast (rapidly fading $\gtrsim$\,0.5 mag per day in the optical), and faint transients (M $\gtrsim$\,$-$16 at peak), e.g., see \cite{2021arXiv210606820A}. Secondly, the typical protocol for identifying and studying KNe remains, for the most part, rapid follow-up of gravitational-wave (GW) and high-energy gamma-ray burst (GRB) triggers (e.g., \citealt{2021arXiv210606820A,2021ApJ...918...63A}). So far, only the BNS merger GW170817 has revealed a KN counterpart, and even with near-term improvements to GW monitoring networks only a few tens of triggers are anticipated throughout the upcoming decade (see \citealt{2019PASP..131f8004A}).

Current and future wide-field optical and near-infrared surveys have been
recently employed as tools for discovering KNe \citep{2019PASP..131f8004A,2021arXiv210606820A,2021ApJ...918...63A}. The fundamental idea is to capitalize on real-time survey data for serendipitous transient discovery, as opposed to ``triggered" observations that use timing and/or localization information from other wavelengths or messengers \citep{2021ApJ...918...63A}. Active facilities applicable for serendipitous fast-transient discovery include the Panoramic Survey Telescope and Rapid Response System (Pan-STARRS; \citealt{2012SPIE.8444E..0HM}), Asteroid Terrestrial-impact Last Alert System (ATLAS; \citealt{2018PASP..130f4505T}), the Dark Energy Camera (DECam; \citealt{2015AJ....150..150F}), the Zwicky Transient Facility (ZTF; \citealt{2019PASP..131f8003B,2019PASP..131g8001G,2019PASP..131a8003M,2020PASP..132c8001D,Ho2022arXiv220112366H}) and the Gravitational-Wave Optical Transient Observer (GOTO; \citealt{GOTO}), while future instrumentation include BlackGEM \citep{2015ASPC..496..254B} and the Vera C. Rubin Observatory's Legacy Survey of Space and Time (LSST; \citealt{2019ApJ...873..111I}).

The ZTF has been a particular focal point for serendipitous KN discoveries as its volumetric survey speed is sensitive to objects that are faint and fast-fading out to almost 200~Mpc \citep{2021ApJ...918...63A}, and its alert stream design follows closely that envisioned for the LSST \citep{2019PASP..131a8001P}. The ZTF was built upon the existing Palomar $48''$ telescope after being equipped with a custom-built wide-field camera (e.g., \citealt{2019PASP..131f8003B}), and its observing system scans large areas of the sky several times each night in multiple bands ($g^\prime r^\prime i^\prime $), with optical transients identified in near-real time using reference image subtraction \citep{2020PASP..132c8001D}.

A first initiative \citep{2021ApJ...918...63A} was developed as the ZTF REaltime Search and Triggering (ZTFReST) to identify KNe in ZTF data, and autonomously rank candidates in the ZTF alert stream based on their photometric evolution and fitting to KN models. Though ZTFReST proved to be effective at serendipitously discovering extragalactic fast transients in the initial system-test trials carried out by \cite{2021ApJ...918...63A}, no KNe were identified. Such results, in addition to outcomes from simulation studies of Vera C. Rubin Observatory observations (e.g., \citealt{2021arXiv210606820A}), have raised the necessity of exploring alternate strategies for synoptic survey techniques and alert characterizations that are optimized to study the transient universe (e.g., see \citealt{2019PASP..131f8004A,2020MNRAS.495.4366A}).

In this article, we propose a method to explore the detection and early characterization of potential KN
candidates from the public data released by ZTF in real time. We use \fink\footnote{\url{https://fink-broker.org}} \citep{10.1093/mnras/staa3602}, a community broker for the upcoming LSST, 
which currently analyzes 
the public alert stream from the ZTF survey. We have built three specific selection algorithms (known as ``filters'') to select the most promising KN
candidates from the global alert streams. The KN 
filters are based on temporal light curve and color evolution, and will be explained in detail in Section \ref{subsec:knfilters}.

We demonstrate the ability of the GRANDMA \citep{GRANDMAO3A,GRANDMA03B} world-wide network of telescopes to respond quickly to \ztfink{} alerts and provide complementary observations to ZTF data at very early times. From the actual cadence of ZTF, we expect a collection of consecutive points, taken with the same filter, to have $\sim2.5$ days latency. However, works such as \cite{implic2020,Almualla_2021} demonstrate that there is added value in obtaining a more refined sampling resolution for the light curve of the transient due to the fast-fading nature of KNe.
Early-time observations are especially important in bluer bands such a $B$ or $g^\prime$, in which the
KN
is expected to fade even more quickly -- becoming undetectable within on the order of one or two days. In addition, incorporating different filter combinations can help in distinguishing KNe
from other candidate transients through their expected color evolution \citep{margutti2018target,CoVi2019,implic2020}. 

The main challenge is not only to coordinate responses to alerts within 48 hours globally, but also to build a real-time data reduction pipeline that is able to digest heterogeneous data from a diverse set of telescopes in order to produce refined data in between consecutive ZTF observations. Although this general problem of calibration, explored for example in \cite{2019MNRAS.484.1031P,2022arXiv220102635B}, is not new, it is crucial for the progress of transient and GW science. The larger objective is therefore to efficiently characterize all of the candidate transients through the use of optimal filter choices and multiple early-time observations, in order to rapidly rule out (or confirm) the nature of the transient as a KN.

We organized an observational campaign from 21 May 2021 to 21 September 2021, named ``ReadyforO4'' (RO4), to invite members of GRANDMA (both professional and amateur astronomers) to follow-up KN candidates from \ztfink{} in order to enable the early characterization of the detected transients. The paper is organized as follows: Section \ref{tel} introduces the GRANDMA consortium, and its citizen science program, Kilonova-catcher. Section \ref{finkKn} provides details on the KN filters used in this work, provided by \fink\ in collaboration with GRANDMA. Section \ref{readyforO4} presents the GRANDMA observations of the RO4 campaign, general data reduction, accuracy of real-time scoring and the results for non-confirmation of KNe. We finish by presenting our conclusions in Section \ref{conclusions}.

\section{GRANDMA and Kilonova catcher}
\label{tel}

The GRANDMA (Global Advanced Rapid Network Devoted to Multi-messenger Addicts) consortium is a world-wide network of 30 telescopes from 23 observatories, 42 institutions, and groups from 18 countries (e.g., see \citealt{GRANDMAO3A,GRANDMA03B}). These facilities make available large amounts of observing time that can be allocated for photometric and/or spectroscopic follow-up of transients. The network has access to wide field-of-view
telescopes ([FoV] $>1$ deg$^2$) located on three continents, and remote and robotic telescopes with narrower fields-of-view
as reported in Table~\ref{tab:GRANDMAtelphoto}.

New telescopes have joined the network since the third LIGO-Virgo observational run O3 (which was described in \citealt{GRANDMAO3A,GRANDMA03B}). In particular, the collaboration with Thailand has provided an opportunity to access both the Southern and Northern sky by Thai Robotic Telescopes (TRTs) located at Springbrook Observatory (TRT-SBO) in Australia and Sierra Remote Observatory (TRT-SRO) in USA.

The 50cm Ali telescope of the Beijing Planetarium, Beijing Academy of Science and Technology, performs both astronomy research and public outreach functions. It is located in the Ngari region of Tibet. It is a $20''$ reflecting telescope equipped with a $22.5\textnormal{mm}\times22.5\textnormal{mm}$ CMOS camera. The $5\sigma$ magnitude limits are expected to be $g^\prime>20$ mag in single images.

A 38cm Schmidt-Cassegrain (with f/11 equipped with a ST-8 CCD sensor) was used at the Lisnyky Observatory granted by the Main Astronomical Observatory of National Academy of Sciences of Ukraine.

The ICAMER Observatory of NAS of Ukraine provided observations with the 60cm Zeiss at the Terskol Observatory located in the North Caucasus.

In addition, the 1m telescope in Pic du Midi is being renovated by the IRAP laboratory and will participate in the GRANDMA consortium, as well as the 1.6m Perkin-Elmer telescope located at Pico dos Dias Observatory in Brazil. All characteristics of the new partners can be found in Table~\ref{tab:GRANDMAtelphoto}.

Within 24 hours, the GRANDMA network is able to access more than $\sim72$\% of the sky (up to more than 80\%) to a limiting magnitude of $\sim18$ mag AB. Due to a less dense distribution of Western Hemisphere observatories, the sky coverage is reduced to $49-60$\% during the night in the Americas. GRANDMA has access to four spectroscopic instruments with sensitivity down to $\approx22$ mag as shown in Table~\ref{tab:GRANDMAtelspectro}. In particular, the High-Energy Transients and their Hosts (HETH) group at the Instituto de Astrof\'isica de Andaluc\'ia has access to two telescopes via competitive proposals for orphan KNe. At the Centro Astron\'omico Hispano en Andaluc\'ia (CAHA), Almeria, Spain, we obtained observing time (PI: Kann) at the 2.2m telescope (equipped with the CAFOS and BUSCA optical imagers) and the 3.5m telescope (with the $\Omega2000$ infrared imager). Two nights of $\Omega2000$ time were granted. At the 10.4m Gran Telescopio Canarias (GTC), we obtained observing time (PI: Kann) with three instruments: the optical imager and spectrograph OSIRIS (8 hrs), the infrared imager and spectrograph EMIR (5 hrs), and finally the integral-field-unit spectrograph MEGARA (2 hrs) to obtain late-time 3D spectroscopy of KN host galaxies. As this is competitive time, we were more conservative with triggering; in particular the GTC proposal was focused on obtaining data from confirmed KNe only. Therefore no observations were obtained during this campaign. Finally, GRANDMA obtained 6 h time allocated by CFHT/WIRCAM (using NIR $JHK_S$ filters) in 2021A and 2021B that were not used due to the lack of confirmed KNe.\\

\begin{figure*}
\begin{center}
\begin{minipage}{0.49\linewidth}
\includegraphics[trim=150 150  150 150,clip=true,width=1.0\textwidth]{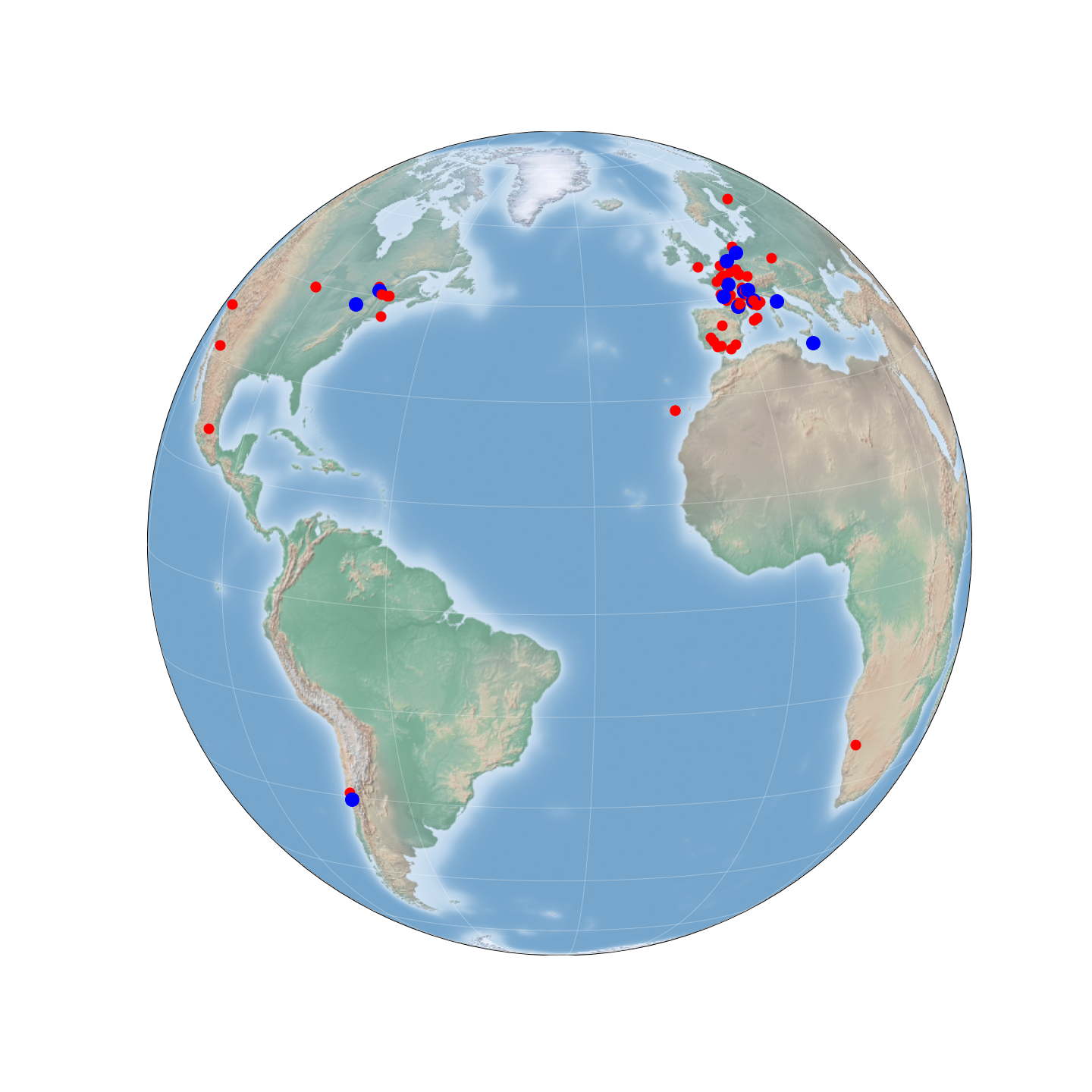}
\end{minipage}
\begin{minipage}{0.49\linewidth}
\includegraphics[trim=150 150  150 150,clip=true,width=1.0\textwidth]{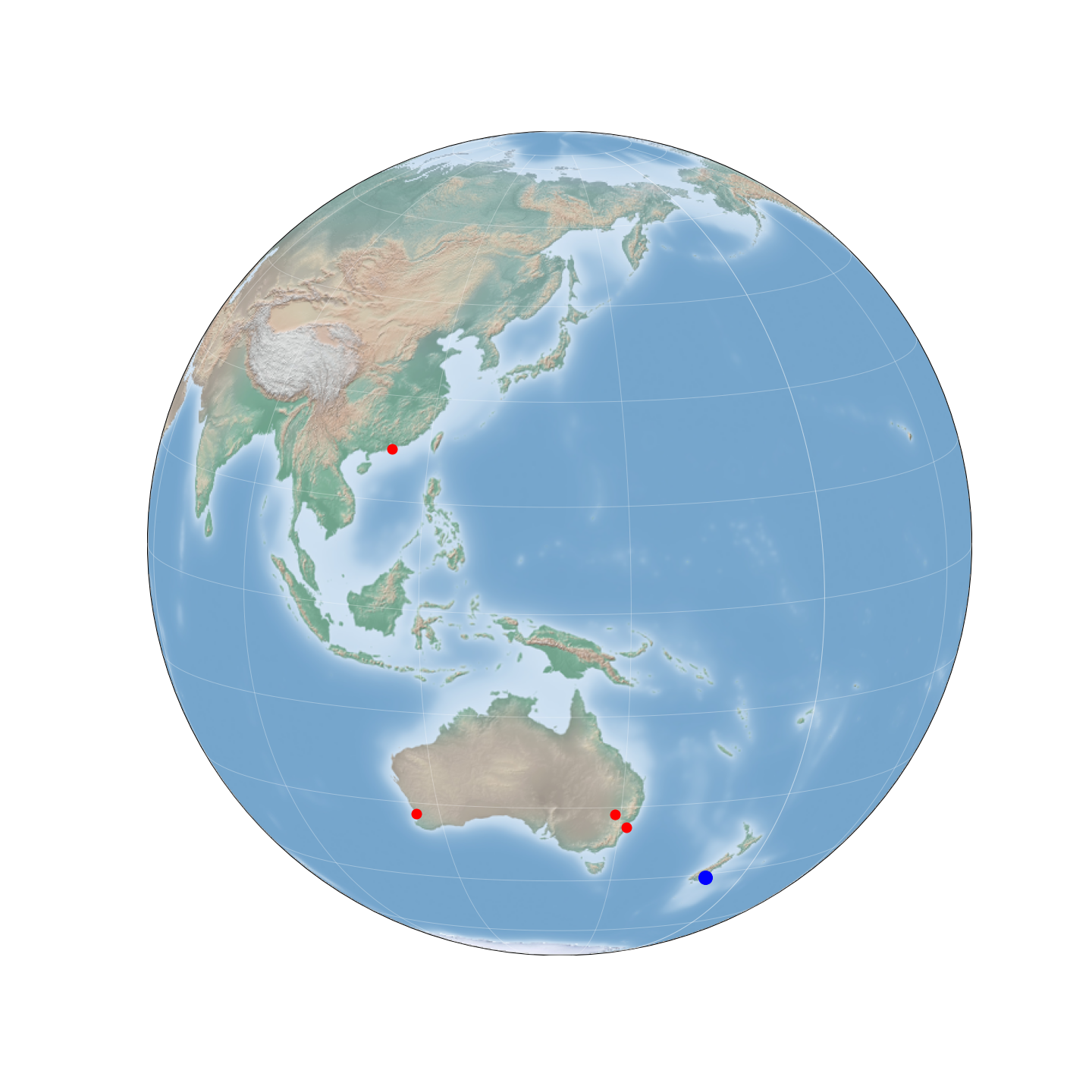}
\end{minipage}
\caption{Locations of the 77 telescopes involved in the GRANDMA citizen science program \textit{Kilonova-Catcher}. In blue are the telescopes used in this study, in red, the rest of the netowrk.} 
\label{fig:KN_user_map}
\end{center}
\end{figure*}

In 2019, GRANDMA initiated the creation of an innovative citizen science program called \textit{Kilonova-Catcher}, hereafter KNC \citep{GRANDMAO3A}. It aims at incorporating amateur astronomers into the search for and follow-up of fast transients such as GRBs and KNe. The GRANDMA consortium has already demonstrated its ability to forward the alerts of the O3 observing run to the amateur astronomers and to provide customized observation plans to each KNC astronomer; in return, they transfer images to the GRANDMA server to be analyzed afterwards. This process creates a continuous chain of observations at very early times after the alerts, before passing the responsibilities to the larger-aperture telescopes of the professional community. KNC uses a dedicated portal \footnote{\url{http://kilonovacatcher.in2p3.fr/} supported by the University of Paris and the IJCLab institute} to organize its activities, while the GRANDMA consortium is in charge of the data analysis of the images (see Section~\ref{readyforO4}). In Fig. \ref{fig:KN_user_map}, we show the locations of the KNC telescope network.

\begin{table*}
\caption{List of telescopes of the GRANDMA consortium and their photometric performance when using their standard setup. In \textcolor{blue}{blue} are mentioned the ones that were engaged in observations for this work.}
\begin{tabular}{ccccccc}
\hline
\hline
Telescope      & Location        & Aperture & FOV            & Filters                                 & $3\sigma$ limit     & Max. Night slot\\
Name           &                 & (m)      &  (deg)     &                                         & (AB mag)             & (UTC) \\
\hline
\textcolor{blue}{TRT-SBO}      & Sierra Remote Obs.   & 0.70 & $0.17\times0.17$   & $UBVR_CI_C$    &  19.0 in 60s (Clear)           & 00h-10h\\
TAROT/TCH      & La Silla Obs.   & 0.25 & $1.85\times1.85$   & Clear, $g^\prime r^\prime i^\prime $    &  18.0 in 60s (Clear)           & 00h-10h\\
\textcolor{blue}{TRT-SRO}      & Springbrook Research Obs.   & 0.70 & $0.17\times0.17$   & $UBVR_CI_C$    &  19.0 in 60s (Clear)           & 10h-16h\\
CFHT/WIRCAM    &    CFH Obs.             &    3.6  &      $0.35\times0.35$             &        $JHK_S$                                &      22.0 in 200s ($J$)       & 10h-16h\\
\textcolor{blue}{FRAM-Auger}     &   Pierre Auger Obs. &  0.30 &  $1.0\times1.0$  & $BVR_CI_C$, Clear   & 17.0 in 120s ($R_C$)          & 00h-10h\\
CFHT/MEGACAM   &     CFH Obs.            &     3.6  &      $1.0\times1.0$               &  $g^\prime r^\prime i^\prime z^\prime$                                       &         23.0 in 200s ($r^\prime$ )    &  10h-16h \\
Thai National Tel. & Thai National Obs.  & 2.40 & $0.13\times0.13$ & Clear, $u^\prime g^\prime r^\prime i^\prime z^\prime$ & 22.3 in 3s ($g^\prime$) & 11h-23h \\
Zadko          & Gingin Obs.      & 1.00 & $0.17\times0.12$ & Clear, $g^\prime r^\prime i^\prime I_C$ & 20.5 in 40s (Clear)  & 12h-22h\\
TNT            & Xinglong Obs.   & 0.80 &  $0.19\times0.19$  & $BVg^\prime r^\prime i^\prime$          &    19.0 in 300s ($R_C$)  & 12h-22h\\
Xinglong-2.16  & Xinglong Obs.   &  2.16    &      $0.15\times0.15$              &     $BVRI$                                    &      21.0 in 100s ($R_C$)       & 12h-22h \\
GMG-2.4        & Lijiang Obs.    &  2.4    &     $0.17\times0.17$                &         $BVRI$                                &     22.0 in 100s ($R_C$)        & 12h-22h\\
\textcolor{blue}{BJP/ALi-50 }      & ALi Obs.    &  0.5    &     $0.38\times0.38$                &         $Clear,g'r'$                                &     20.8 in 20s (Clear)        & 14h-00h\\

\textcolor{blue}{UBAI/NT-60}     & Maidanak Obs.   & 0.60 & $0.21\times0.21$   & $BVR_CI_C$                              &    18.0 in 180s ($R_C$) & 14h-00h\\
UBAI/ST-60     & Maidanak Obs.   & 0.60 & $0.23\times0.23$   & $BVR_CI_C$                              &    18.0 in 180s ($R_C$) & 14h-00h\\
TAROT/TRE      & La Reunion      & 0.18 &  $4.2\times4.2$  & Clear                                   &     16.0 in 60s (Clear)        & 15h-01h\\
Les Makes/T60  & La Reunion.     & 0.60 &  $0.3\times0.3$  & Clear   &   19.0 in 180s (Clear)       & 15h-01h\\
\textcolor{blue}{Terskol/Zeiss-600} & Terskol Obs & 0.6 & 0.18x0.18 & Clear, $BVR_CI_C$ & 21.5 in 120s ($R_C$) & 16h-02h \\
\textcolor{blue}{Abastumani/T70} & Abastumani Obs. & 0.70 & $0.5\times0.5$   & $BVR_CI_C$                              &    18.2 in 60s ($R_C$)   & 17h-03h\\
ShAO/T60       & Shamakhy Obs.   & 0.60 & $0.28\times0.28$ & $BVR_CI_C$                              &    19.0 in 300s ($R_C$)  & 17h-03h \\
Lisnyky/AZT-8 & Kyiv Obs.       & 0.70 & $0.38\times0.38$ & $UBVR_CI_C$                             & 20.0 in 300s($R_C$) & 17h-03h\\
\textcolor{blue}{Lisnyky/Schmidt}  & Kyiv Obs.       & 0.36 & $0.20\times0.14$ & Clear, $UBVR_CI_C$  & 19.5 in 300s($R_C$) & 17h-03h\\
TAROT/TCA      & Calern Obs.     & 0.25 & $1.85\times1.85$ & Clear, $g^\prime r^\prime i^\prime$     &  18.0 in 60s (Clear)          & 20h-06h\\
\textcolor{blue}{FRAM-CTA}    & ORM  & 0.25 &  $0.43\times0.43$   & Clear, $BVR_Cz^\prime$,    & 16.5 in 120s ($R_C$)          & 20h-06h\\
\textcolor{blue}{IRIS}           & OHP             & 0.50  &  $0.4\times0.4$  &   Clear, $u^\prime g^\prime r^\prime i^\prime z^\prime $ & 18.5 in 60s ($r^\prime$)   & 20h-06h\\
T120           & OHP             & 1.20 &   $0.3\times0.3$     &     $BVR_CI_C$                                    &   20.0 in 60s ($R$)           & 20h-06h\\
Pic du Midi/T1M           & Pic du Midi & 1.05  &  $0.13\times0.13$  &   $u^\prime g^\prime r^\prime i^\prime z^\prime $ & 19.5 in 60s ($r^\prime$)   & 20h-06h\\
2.2m CAHA/CAFOS      & Calar Alto Obs. & 2.20 &  $0.27\diameter$                & $ u^\prime g^\prime r^\prime i^\prime z^\prime $                              & 23.7 in 100s ($r^\prime$) & 20h-06h\\
3.5m CAHA/$\Omega2000$      & Calar Alto Obs. & 3.50 &  $0.257\times0.257$                & $ JHK_S $                              & 20 in 90s ($J$) & 20h-06h\\
10.4m GTC/OSIRIS      & ORM. & 10.40 &  $0.13\times0.13$                & $u^\prime g^\prime r^\prime i^\prime z^\prime$                              & 24 in 30s ($r^\prime$) & 20h-06h\\
10.4m GTC/EMIR      & ORM. & 10.40 &  $0.111\times0.111$                & $YJHK_S$                              & 24 in 120s ($Y$) & 20h-06h\\
\textcolor{blue}{VIRT}           & Etelman Obs.   & 0.50 & $0.27\times0.27$ & $UBVR_CI_C$, Clear &  19.0 in 120s (Clear) & 22h-04h\\
Perkin-Elmer Tel.  & Pico dos Dias Obs & 1.6   &       $0.083\times0.083$           &  $UBVR_CI_C$ & 21 in 360s(Clear) & 18h-01h \\
\hline
\end{tabular}
\label{tab:GRANDMAtelphoto}
\end{table*}

\begin{table*}
\caption{List of telescopes of the GRANDMA consortium with spectroscopic capabilities.}
\begin{tabular}{ccccccc}
\hline
\hline
Telescope/Instrument & Location & Wavelength range & Spectral resolution $\lambda/\Delta\lambda$ &  Limiting mag\\
\hline
2.2m CAHA/CAFOS   & Calar Alto Obs. & 3200-7000/6300-11000 & 400 & 20 in 1h \\
ShAO/T2m & Shamakhy Obs. & $3800-8000$ & 2000 & 17 in 1h \\
Terskol-2m/MMCS & Terskol Obs. & $3800-9000$ & 1200 & 17 in 1h \\
Xinglong-2.16/BFOSC & Xinglong Obs. & $3600-9600$ & 1000 & 18 in 1h \\
GMG-2.4/YFOSC & Lijiang Obs. & $3400-9100$ & 2000 & 19 in 1h \\
10.4m GTC/OSIRIS & ORM & $3630-7500/7330-10000$ & 1018/2503 & 24 in 1h  \\
10.4m GTC/EMIR & ORM & $890-13310$ & 987 & 21 in 1h  \\
\hline
\end{tabular}
\label{tab:GRANDMAtelspectro}
\end{table*}

\section{Method: searching for Kilonovae using the \fink\ broker}
\label{finkKn}

\subsection{\fink\ overview}

\fink\ \citep{10.1093/mnras/staa3602} is a community broker designed to enable science with large time-domain alert streams such as the one from the current ZTF survey or the upcoming LSST. Driven by its scientific community, \fink\ probes a large number of topics in the transient sky, from solar system science to galactic, and extra-galactic science. 

\fink\ currently analyzes the public alert data stream from the ZTF survey. Each night, alerts are collected in real-time after their processing by ZTF. \fink\ received 35,387,098 alerts between 01 April and 30 September 2021 (160 observing nights). Alerts carry out basic measurements for the trigger (position, magnitude, time), but also the detections that occurred up to 30 days before the alerts at the same location in the sky, allowing to reconstruct at least a partial light curve of the potential candidate. All incoming alerts are stored on disk, but only alerts with sufficient quality are then processed. At the time of the campaign, there were three quality cuts in \fink\ to assess the quality of alerts and reject artefacts and known bogus alerts \citep{10.1093/mnras/staa3602}; these cuts discarded about $70\%$ of the incoming alerts.

\begin{figure}
\begin{center}
\includegraphics[width=0.46\textwidth]{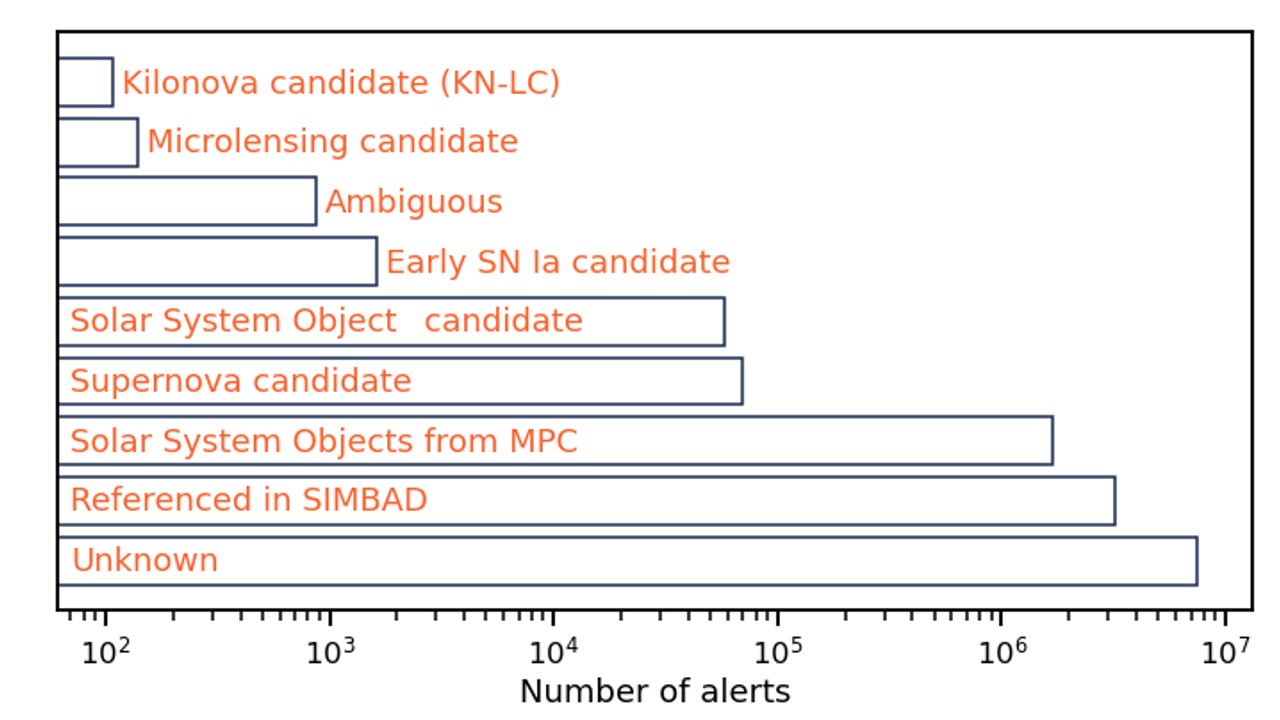}
\caption{\fink\ classification labels for the 12,556,539 ZTF alerts that passed the quality cuts in the period 01 April to 30 September, 2021. About half of the alerts got no classification (Unknown label), i.e., \fink\ was not able to conclude on the nature of the alert given the alert information available. The remaining half is dominated by objects with a counterpart in the SIMBAD database (match within 1\farcs5 radius), and alerts associated with a known object from the Minor Planet Center database (moving objects from the Solar System). Other alerts are associated with supernova events, Solar System Object candidates, microlensing candidate events, or have an ambiguous classification (more than one label at a time). Concerning KN candidates, we report here the 107 alert candidates from the KN-LC filter (see text below).}
\label{fig:fink-classes}
\end{center}
\end{figure}

The remaining $30\%$ of alerts (12,556,539 alerts) are then processed by the \fink\ science modules. Science modules are provided by the community of users to add value to alerts, as detailed below. We note that the state of the broker constantly evolves over time (additions, or corrections), and we report results using \texttt{fink-broker} version 1.1, \texttt{fink-science} version 0.4, and \texttt{fink-filters} version 0.2. There are several types of added values: labels from the cross-match with external catalogs or survey feeds, classification scores provided by a machine learning analysis, or simply tags based on the alert content. These added values are then combined to provide a unique classification for each alert. Fig. \ref{fig:fink-classes} shows the alert classification during the KNC observational campaign. About half of the alerts got classified, i.e., \fink\ can extract information about the potential nature of the object. Most of the classified alerts are cataloged in the SIMBAD database\footnote{\url{https://www.minorplanetcenter.net/}}. For the KN classification, we describe and report below the candidates from the KN-LC filter. 






\subsection{Kilonova candidate selection}
\label{subsec:knfilters}
Our main goal is to identify the most probable KN candidates among all incoming alerts. There are two competing factors: criteria that are too broad would yield too many candidates to follow up, given the huge number of incoming alerts; on the other hand, complex selection criteria would be meaningless given the lack of actual KN observations to constrain the parameter space. 

In order to optimize our search for KNe, we designed three selection filters targeting different likely aspects of a KN. They act at the end of the \fink\ processing to reduce the incoming data stream and to select the most probable KN candidates. In the period of 01 April to 30 September  2021 we obtained:

\begin{itemize}
    \item Machine learning-based filter (KN-LC): 107 alert candidates
    \item Near-by Galaxy Catalogs-based filter (KN-Mangrove): 68 alert candidates
    \item Rate-based filter (KN-Slope): 127 alert candidates
\end{itemize}

We note that only five alert candidates were selected by more than one filter, showing that the filters are triggered by different parameters of the incoming alerts. Hence, alerts selected by more than one filter are particularly noteworthy. All filters are open-source and can be found at \url{https://github.com/astrolabsoftware/fink-filters}. A detailed description and analysis of the filter outputs can be found in the accompanying notebook\footnote{\url{https://github.com/astrolabsoftware/fink_grandma_kn}}.

\subsubsection{Machine learning-based filter (KN-LC)} \label{sec:kn-lc}

This filter mainly uses information from the light curve. During the classification step, the \fink\ classifier extracts features from the light curves in the $g^\prime$ and $r^\prime$ photometric bands (see Biswas et al., in prep.) and infer the probability of an alert being a KN. The light curves are deconstructed as a linear combination of principal components, and additional features are also extracted such as the maximum of the flux, residuals, and the number of measurements. Because we have better accuracy with more principal components, choosing the number of principal components is a compromise between the efficacy of the classifier and the amount of time that is needed to classify alerts and identify candidates (the more components, the more measurements we need, see Biswas et al., in prep. for more details). 

In the early days of the campaign, we were using only the first principal component.
However, it was shown quickly that one component was not sufficient to produce a reliable score on the alert data, and we had many false-positives (e.g., many candidates were obvious supernovae). So we introduced the second principal component in the set of features for the classification, and the results with real data improved without introducing further delays in practice given the cadence of the ZTF survey (that is without the need to add more measurements). In addition, the average number of candidates per month was reduced, from about 28 candidates per month to about 10 per month. The change of the model in the classifier happened on 17 June, 2021. In \fink, data and model are versioned, and the new model corresponds to the version 0.4.5.

\begin{figure}
\begin{center}
\includegraphics[width=0.5\textwidth]{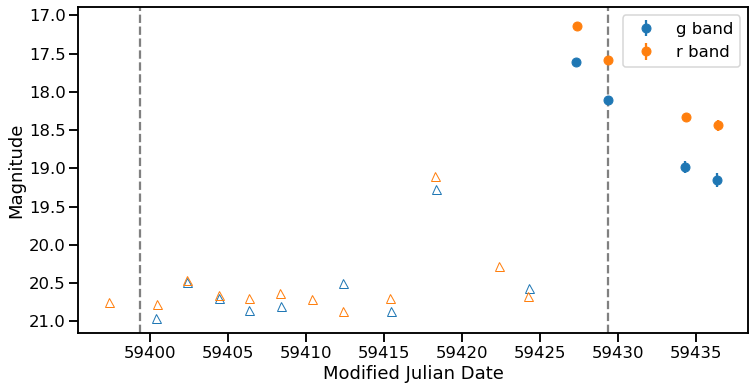}
\caption{ZTF object light curve of one kilonova candidate from the KN-LC filter. Circles with error bars show valid alerts (detections) that pass the \fink\ quality cuts. Downward-pointing open triangles represent $5\sigma$ magnitude limits in template-subtracted difference images based on point-spread-function (PSF) fit photometry contained in the history of valid alerts. The right-most vertical line shows the KN trigger by \fink, and data used to classify the alerts are shown in between dashed vertical lines (30-day-history data are attached with the alert). The alert leading to the KN trigger, \href{https://fink-portal.org/ZTF21abqfzcp}{ZTF21abqfzcp}, was emitted on 2021-08-03 08:51:48 UTC. On the next re-observing night (five days after), new photometric data from alerts favoured a supernova candidate classification, ruling out the nature of the transient being a KN.}
\label{fig:ztf21abvrrir}
\end{center}
\end{figure}

This filter uses the five following criteria:
\begin{itemize}
    \item The score from the KN classifier must be above 0.5 (binary random forest classifier).
    \item Point-like object: the star/galaxy extractor score must be above 0.4.
    \item Non-artifact: the deep real/bogus score must be above 0.5.
    \item Object not referenced in the SIMBAD database (except from extra-galactic origin).
    \item Young detection: less than 20 days. This threshold is quite loose but it is sufficient to filter long-trend or well-known objects.
\end{itemize}

Over the campaign, this filter selected 107 alert candidates out of 12,556,539. This corresponds to 70 unique objects on the sky (the same astrophysical objects can emit several alerts over time). Fig. \ref{fig:ztf21abvrrir} shows the light curve of such a candidate. All objects are labeled in the \fink\ database and can be easily accessed via the Science Portal\footnote{\url{https://fink-portal.org}}, or via the REST API.

With time, \fink\ collects more alert data, and has a clearer view on the nature of each object. At the end of the campaign, we found that most objects that emitted at least one alert tagged as KN candidates were deemed to be supernova candidates; however, some remained as KN candidates. 

We also performed a crossmatch with the data from the Transient Name Server. Considering only the candidates from the first model (before June 2021), most of the candidates turned out to be Type Ia supernovae (41/71). However, after the model used to classify alerts changed, 29/36 of the candidates had no counterpart in the Transient Name Server (i.e., there was no follow-up), the others being identified as cataclysmic variables (6/36) or Type IIb supernovae (1/36).




\subsubsection{Near-by Galaxy Catalogs (KN-Mangrove)} \label{sec:kn-mangrove}

With the previous filter, KN-LC, we concluded that a minimum of two days from the first detection by ZTF is necessary to get a reliable score from the classifier and to identify candidates. According to KN models, two days after the compact binary merger, the signal will be fading or even too faint to be observed. So we developed a second filter to address younger detections.  This filter uses the following criteria:

\begin{itemize}
    \item Point-like object: the star/galaxy extractor score must be above 0.4.
    \item Non-artifact: the deep real/bogus score must be above 0.5.
    \item Object not referenced in the SIMBAD database (except from extra-galactic origin).
    \item Young detection: less than 6 hours.
    \item Galaxy association: the alert should be within 10 kpc of a galaxy from the Mangrove catalog \citep{Berger:2013jza}. The 10 kpc is empirical -- we also tested different values. Above 10 kpc, we have a very large rate of contaminants. Below 10 kpc, we would potentially miss valid transients.
    \item Absolute magnitude: the absolute magnitude of the alert should be $-16\pm1$ mag (in both $g^\prime$ and $r^\prime$ bands).
    \item Non Solar System Object: the alert must be at least 5\arcsec\ away from any known Solar System objects referenced in the Minor Planet Center database at the time of emission.
\end{itemize}

\begin{figure}
\begin{center}
\includegraphics[width=0.5\textwidth]{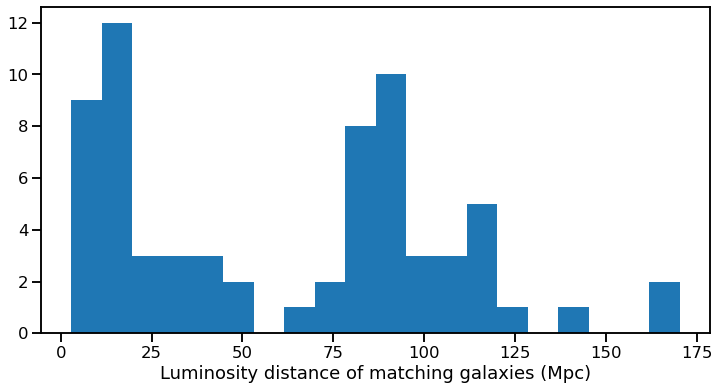}
\caption{Histogram of the luminosity distance of galaxies associated with alerts selected by the KN-Mangrove filter. For the campaign, we only considered the galaxies from the Mangrove catalog in a 230 Mpc range, corresponding to the observation range of current GW detectors. In practice, we have no candidates further than 170 Mpc. Alerts and galaxies are matched within a radius of 10 kpc. The threshold has been set such to avoid a large number of spurious associations in the observation plane, while allowing coverage around the galaxy. Note that the allowed angular distance between alerts and galaxies increases when the luminosity distance decreases, hence leading to more frequent spurious associations at small luminosity distances.}
\label{fig:galaxy-luminosity-distance}
\end{center}
\end{figure}

With the KN-Mangrove filter, an alert will be considered as a candidate if one can identify a suitable host and the resulting absolute magnitude is compatible with KN models. To identify possible hosts, we used the MANGROVE catalog \citep{10.1093/mnras/staa114}, containing about 800,000 near-by galaxies. For the campaign, we only considered galaxies within 230 Mpc, as it corresponds to the current observation range of GW detectors. Fig. \ref{fig:galaxy-luminosity-distance} shows the distribution of luminosity distances of galaxies associated with alerts selected by the KN-Mangrove filter.

In practice, the galaxy association method is not perfect, and can lead to misassociation of an event that is in the foreground or the background of a galaxy\footnote{See for example \url{https://fink-portal.org/ZTF21abdwdwo} that was selected by the KN-Mangrove filter and associated with the bright LEDA 1740743 galaxy in Mangrove. After visual inspection, it turns out that the alert more likely originated from a fainter galaxy located further away, and not present in Mangrove.}. But this is inevitable, as the luminosity distance between the Earth and the source generating the alert is usually unknown.

According to \citet{Kasliwal_2020}, we expect a KN event to have a peak absolute magnitude at $g^\prime \sim-16$ mag. This threshold is given in $g^\prime$-band, but in this work it was implemented for $g^\prime$ and $R^\prime$ bands without distinction. This hypothesis is due to the lack of early observations and strong consistency with AT2017gfo \citep{LSC_MM_2017ApJ}. As we often do not know the source distance, we compute the absolute magnitude from an alert as if it were in the matched galaxy.

68 alert candidates were selected by this filter out of 12,556,539 processed alerts from 01 April to 30 September, 2021. This corresponds to 59 unique objects on the sky.  At the end of the campaign we checked, using more data,  the evolution of the classification of those objects. We found that most objects remained KN candidates according to the KN-Mangrove classifier, while a small fraction were subsequently identified as potential supernova candidates or Solar System Object candidates. This was confirmed when checking against Transient Name Server data, where we found 51/68 alerts without a counterpart (i.e., no follow-up result was reported), 7/68 confirmed as supernova type Ia, 4/68 as supernova type II, 3/68 as supernova type IIp, 1/68 as supernova type IIb, 1/68 as supernova type Ib, and 1/68 as supernova type Ic.

\subsubsection{Slope-based filter (KN-Slope)} \label{sec:kn-slope}

In addition to the two previous filters, we developed a third filter based on the work of \cite{2021ApJ...918...63A}. The main criterion used for extracting KN candidates corresponds to the slope of the normalized light curve (mag/day). Given the expected light curves of KNe, we chose a threshold of 0.3 mag/day, which corresponds to a fast-fading object. This filter was not adopted during the campaign as it did not provide satisfying results. However for reference we reprocessed the campaign data and we present its performance here. This filter will be used in the subsequent campaigns along with the two other filters. In total, this filter implements eight criteria:

\begin{itemize}
    \item Fast fade: The apparent magnitude decay rate of the alert must be above 0.3 mag/day in the last photometric band with two sequential measurements.
    \item Point-like object: The star/galaxy extractor score of the alert must be above 0.4.
    \item Non-artifact: The deep real/bogus score of the alert must be above 0.9.
    \item Object not referenced in the SIMBAD database (except from extra-galactic origin).
    \item Object not referenced in the Sloan Digital Sky Survey (SDSS) as a star or quasi-stellar object.
    \item Young detection: The alert emission date must be less than 14 days. In practice the delay between  the first detection and the trigger by \fink\ is a maximum two days (two consecutive measurements by ZTF).
    \item Non Solar System Object: The alert must be at least 10\arcsec away from any known Solar System objects referenced in the Minor Planet Center database at the time of emission.
    \item Away from the galactic plane: The alert must have an absolute galactic latitude above 10 degrees.
\end{itemize}

127 alert candidates were selected by this filter out of 12,556,539 processed alerts from 01 April to 30 September, 2021. This corresponds to 108 unique objects on the sky. At the end of the campaign, most of the objects had produced more alerts after the initial KN trigger, and were mostly falling under the class of supernova candidates according to the \fink\ filters. When cross-matching with data on the Transient Name Server, 117/127 have no counterparts, 6/127 are from cataclysmic variables, 3/127 are from supernovae type Ia, and 1/127 is from supernova type IIn.

\subsection{Accuracy, efficiency, and added value}

The main purpose of the classifier is to identify the most probable fast transient candidates in the sample, with a particular focus on KNe. However, to determine the nature of transients of interest, it generally requires follow-up photometry above and beyond what is provided by a survey with return timescales of $\sim1$ night or more. For this reason, measurements of the number of transients passing the selection filters are required to understand the nature of objects that will pass those filters. In particular, empirical measurements of the rate of transients passing particular filters are useful for assessing the contamination rate and therefore the follow-up photometry (and potentially spectroscopy) required to characterize the sample.

\begin{figure}
\begin{center}
\includegraphics[width=0.5\textwidth]{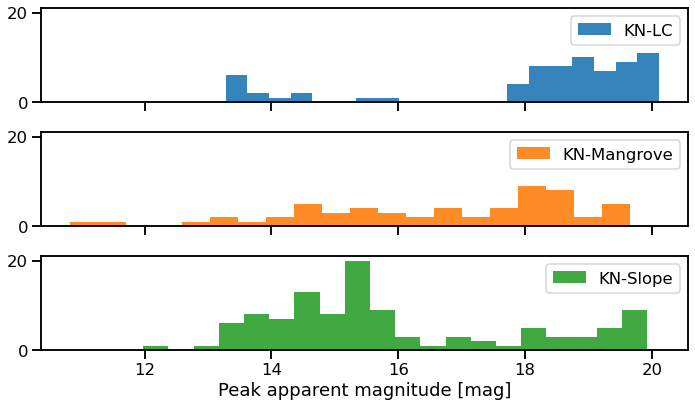}
\caption{Histogram of the peak apparent magnitude for the candidates that pass the three filters: KN-LC (top, blue, see \ref{sec:kn-lc}), KN-Mangrove (middle, orange, see \ref{sec:kn-mangrove}), KN-Slope (bottom, green, see \ref{sec:kn-slope}). In the case of the KN-LC and KN-Slope filters, the distributions have an excess of candidates at the two ends (faintest and brightest objects), while the peak apparent magnitude distribution in the case of KN-Mangrove is rather uniform across the magnitude range. The faintest peak apparent magnitude is around 20 mag for all filters.}
\label{fig:peak-app-mag}
\end{center}
\end{figure}

To evaluate the contaminant rate, we re-analyzed the ZTF alert stream data taken between November 2019 and September 2021 (538 observing nights), corresponding to 38,372,852 (12,149,579) processed alerts (unique objects). Using this data, 197 (132) alert candidates have passed the KN-LC filter, 227 (208) candidates passed the KN-Mangrove filter, and 285 (247) candidates passed the decay rate filter (KN-Slope). According to Transient Name Server classification data, for the KN-LC and KN-Mangrove filters, the most common contaminant transients were supernovae near peak\footnote{Note that in the case of KN-Mangrove, we would have four times more candidates if we did not reject Solar System objects.}, and for the decay rate filter, the most common transients were fading cataclysmic variables near the Galactic plane. Given the ZTF coverage, this corresponds to a magnitude-limited rate of about 0.5 candidates per night per filter down to a magnitude of 20.5. Fig. \ref{fig:peak-app-mag} shows the histogram of the peak apparent magnitude for the candidates that pass the filters. While the KN-LC and KN-Slope distributions seem bi-modal (due to, for example, the excess of cataclysmic variables in the KN-Slope filter), the results from the KN-Mangrove filter are more spread across the magnitude range.  Our analysis provides important information to understand the underlying population of contaminants (assuming there are no KNe) and also for future surveys as Vera Rubin LSST.

\section{GRANDMA/Kilonova-catcher RO4}
\label{readyforO4}

We organized the ``ready for O4 campaign-I'' to a) demonstrate the potential of amateur astronomy in the search of GW counterparts, b) introduce the GRANDMA consortium into the search for KNe, and c) establish the caveats for performing joint photometry with different apertures and filters.  From 21 May, 2021 to 21 September, 2021, we followed six alerts sent by our KN broker implemented in \fink\ (see section~\ref{finkKn}) focusing on alerts sent on Friday. The aim was to invite amateurs to observe the transient during the next 72~h (so that the amateurs would have the weekend for performing observations), to verify how many would respond and with which latency. If no alert passed our thresholds, we provided the observers  recent supernovae (three in our campaign) to be observed for practice purposes. In July 2021, we invited all GRANDMA teams to observe kilonova candidates as well in order to have a larger set of images to test our photometric pipeline on heterogeneous data. Several online tutorials were organized allowing us to enroll a large number of amateurs with their respective observatories.  A preliminary requirement file was released to indicate to the observers how to provide useful data for the campaign. In particular, we were firstly targeting a ``classification mode'' requiring the use of several photometric bands to either confirm the event detection or reject it as a false alarm, and secondly a ``monitoring mode'' by following up the multi-band flux evolution of the transient source.

\begin{figure}
\begin{center}
\includegraphics[width=0.50\textwidth]{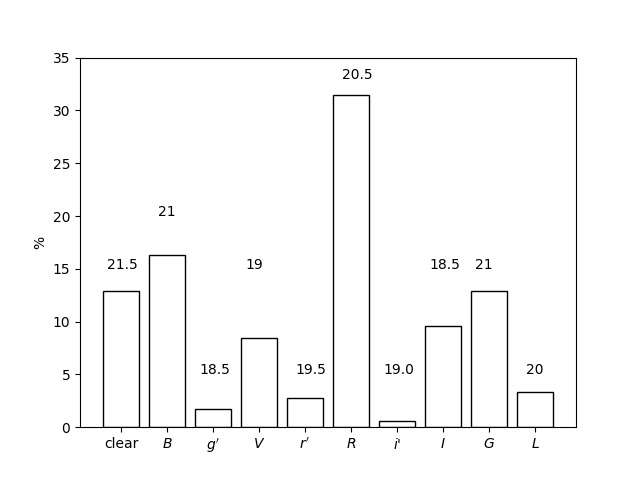}
\caption{Filters used in the images taken during this observational campaign and when a source is detected. Above are mentioned the fainter upper limits reached during the campaign (see section~\ref{datareduc}).}
\label{fig:conclufilters}
\end{center}
\end{figure}

In total, we achieved participation by 26 amateurs and 11 distinct GRANDMA telescopes (see Table~\ref{tab:GRANDMAtelspectro}). We received received images taken with filters in professional and amateur filter systems, and also images taken without filters (see Figure~\ref{fig:conclufilters}). 

\subsection{Data reduction}
\label{datareduc}

In order to uniformly process the diverse set of images acquired by various telescopes, we developed two dedicated photometric pipelines: {\sc STDPipe}\footnote{{{\sc STDPipe} is available at \url{https://gitlab.in2p3.fr/icare/stdpipe}}} (will be referenced henceforth as ``STD'') and {\sc MUphoten}\footnote{{{\sc MUphoten} is available at \url{https://gitlab.in2p3.fr/icare/MUPHOTEN}}} \citep{muphoten} (will be referenced as ``MU'').

\textbf{STDPipe} --
{\sc STDPipe} \citep{stdpipe} is a set of Python routines for astrometry, photometry and transient detection related tasks, intended for quick and easy implementation of custom pipelines, as well as for interactive data analysis. It is designed to operate on standard Python objects: NumPy arrays for images, Astropy Tables for catalogs and object lists, etc., and conveniently wraps external codes that do not have their own Python interfaces (SExtractor \citep{sextractor}, SCAMP \citep{scamp}, PSFEx \citep{2011psfex}, HOTPANTS \citep{hotpants}, Astrometry.Net \citep{2010astrometrynet}, etc). It supports the following steps of processing and analyzing the images:

\begin{itemize}
\item object detection and photometry using either SExtractor \citep{sextractor} or SEP \citep{sep} codes. Simple PSF photometry may be performed by the SExtractor backend using a PSF model estimated by the PSFEx \citep{psfex} software, or aperture photometry based on photutils \citep{photutils} may be run using various kinds of background estimation, either global or local.
\item astrometric calibration using Astrometry.net \citep{2010astrometrynet} for blind World Coordinate System (WCS) solving in either remote or locally-installed variants, and using SCAMP \citep{scamp} or custom Astropy-based code for astrometric refinement by matching the lists of objects detected in image with catalog entries
\item photometric calibration using any catalog available in Vizier as a reference with approximate on-the-fly passband conversion for some of them (in order to derive the magnitudes in the Johnson-Cousins system based on the ones in Pan-STARRS or Gaia systems). A sophisticated photometric matching routine is available for fitting the zero point and photometric system of the frame taking into account a spatial polynomial of arbitrary order, a color term, as well as an optional additive flux term.
\item image subtraction with the HOTPANTS \citep{hotpants} code using either locally available templates or images automatically downloaded from the network.
The code allows downloading templates either from the Pan-STARRS archive of stacked images \citep{ps1images}, or from any imaging survey accessible through the HiPS2FITS \citep{hips2fits} service. When running the image subtraction code, a custom noise model may be supplied in order to account for the poorly known gain and bias levels of the image.
\item transient detection and photometry on difference images taking into account the proper noise model of the difference image and various artifact masks in order to decrease the number of false detections. The transients may be filtered based on coincidences with either locally available or remote catalogs, as well as with positions of known Solar System objects. Also, optionally a routine for sub-pixel adjustment of the transient cutout and template images may be performed in order to detect cases of slight positional shifts (causing ``dipoles'' in difference images) either due to overall image misalignment, or object proper motion due to a large difference in the template epoch. 
\item insertion of simulated stars with realistic PSFs into the images in order to assess the performance of transient detection.
\item the code also includes various convenience utilities and plotting routines for quick visualization of the results of every step.
\end{itemize}

The actual image processing pipeline for the present work was organized as follows.
As {\sc STDPipe} is intended for higher-level analysis of pre-processed frames, we required all images to be pre-processed by an instrument-specific code to perform bias, dark subtraction, and flat-fielding in advance.
Then we removed the cosmic rays using the astroscrappy \citep{astroscrappy} code implementing the original LACosmic \citep{lacosmic} algorithm, and detected the objects on the image using SExtractor \citep{sextractor}. Next, we performed aperture photometry with local background subtraction using a photometric aperture with radius equal to the median image full width at half maximum (FWHM), and a background annulus between radii of 5 and 7 FWHM units. Then, the astrometric solution was derived using the
Astrometry.net \citep{2010astrometrynet} solver applied to the list of detected objects. Then, we downloaded the list of stars from the Pan-STARRS DR1 catalog from Vizier to serve as both an astrometric and photometric reference catalog, and augmented it with Johnson-Cousins $B$, $V$, $R_C$ and $I_C$ magnitudes using approximate conversions derived by \citet{ps1_to_stetson}.
We refined the image astrometric solution using the SCAMP \citep{scamp} code, via lists of detected objects and catalog stars. Then we constructed the photometric solution for the image using the closest Johnson-Cousins filter as a reference and $B-V$ (or corresponding Pan-STARRS filter and $g^\prime-r^\prime$ for the telescopes using Sloan-like filter sets) as a color used for deriving an instrumental photometric system (color term). For the zero point, we used either a constant value for all stars if the field of view (FOV) and number of stars were small, or a second order spatial polynomial if there were enough stars in the frame. Then we downloaded the Pan-STARRS co-added images covering the observed field of view in the closest filters (either $g^\prime$, $r^\prime$ or $i^\prime$), mosaiced them and used the result as a template which was then subtracted from the image using the HOTPANTS \citep{hotpants} code. On the resulting difference image, we performed forced aperture photometry at the transient position using the same settings as used for deriving the original photometry, so that the zero point model was still valid, and thus derived the transient photometry in a system linked to a standard one (either Johnson-Cousins or Pan-STARRS) by a color term value specific for this frame.
In a similar manner, we determined an effective detection limit at the transient position by converting the background noise inside the aperture multiplied by 5 (so that it corresponds to 5$\sigma$) to flux and then to the magnitude. When the object is not detectable in the image, this value was adopted as an upper (detection) limit for its magnitude. 

\textbf{MUphoten} --
{\sc MUphoten} \citep{muphoten} is a Python-based software dedicated to photometry of transients followed up by heterogeneous telescopes. It uses public Python libraries such as Photutils \citep{photutils}, Astroquery \citep{astroquery}, and also uses external C codes: SExtractor \citep{sextractor}, Scamp \citep{scamp}, Swarp \citep{2002ASPC..281..228B} and HOTPANTS \citep{hotpants}. The pipeline works on pre-processed images (dark or bias subtracted, flat-fielded) and for which an astrometric solution is known. For this campaign the astrometric calibration was done directly on the Astrometry.net website \citep{2010astrometrynet}. The analysis process works as follows:
\begin{itemize}
\item subtraction of a Pan-STARRS template constructed with a mosaic of the observed FOV downloaded from the catalog archives. For non-Sloan filters (e.g., Johnson-Cousins or unfiltered images), the closest band of the Pan-STARRS system was used: $g_{ps1}$ for $B$, $V$ and clear images, $r_{ps1}$ for $R_C$ images, and $i_{ps1}$ for $I_C$.
\item background estimation in a mesh of $150\times150$ pixels, using the same estimator as SExtractor. The background is then subtracted from the image. 
\item source detection using a $2\sigma$ threshold above the background.
\item aperture photometry on the detected sources.
\item cross-match with Pan-STARRS catalog to fit the instrumental magnitude versus Pan-STARRS magnitude relation with a linear fit. For images acquired using filters of the Johnson-Cousins photometric system, we used the second order equations 1, 2, 4 and 6 of Table 2 from \citet{ps1_to_stetson} to transform the Pan-STARRS system to the observed filter. For clear images, we added the flux from the $g_{ps1}$ and $r_{ps1}$ Pan-STARRS bands.
\item detection of the transient in the residual image between the observed and the template images, adopting a $3\sigma$ threshold. A positive detection of a transient was considered if a source was identified at less than five pixels from the transient position.
\item evaluation of the transient instrumental magnitude in the residual image between the observed and the template image from the Pan-STARRS survey and usage of the previous fit to obtain the calibrated measurement of the transient.
\end{itemize}


{\sc MUphoten} uses two methods to filter out poor quality images. One by comparing the calibrated magnitude of a star in the field of view to its magnitude in the catalog used for the photometric calibration. If they are incompatible, the image is rejected. 
The second veto consists in computing the seeing of the image with PSFex \citep{2011psfex}. Then for a given band of a given telescope+filter configuration, reject the images for which the seeing deviates by more than $3\sigma$ from the median seeing of the data set. For the images passing the vetoes with no detected transient, we set an upper limit on the magnitude. It is estimated by dividing the number of detected objects in the image by the number of objects in the reference catalog in magnitude bins. When the result drops below 0.5, the center of the corresponding bin is considered to be the limiting magnitude.

\subsection{Consistency of the analysis}

\begin{figure}
\begin{center}
\includegraphics[width=0.45\textwidth]{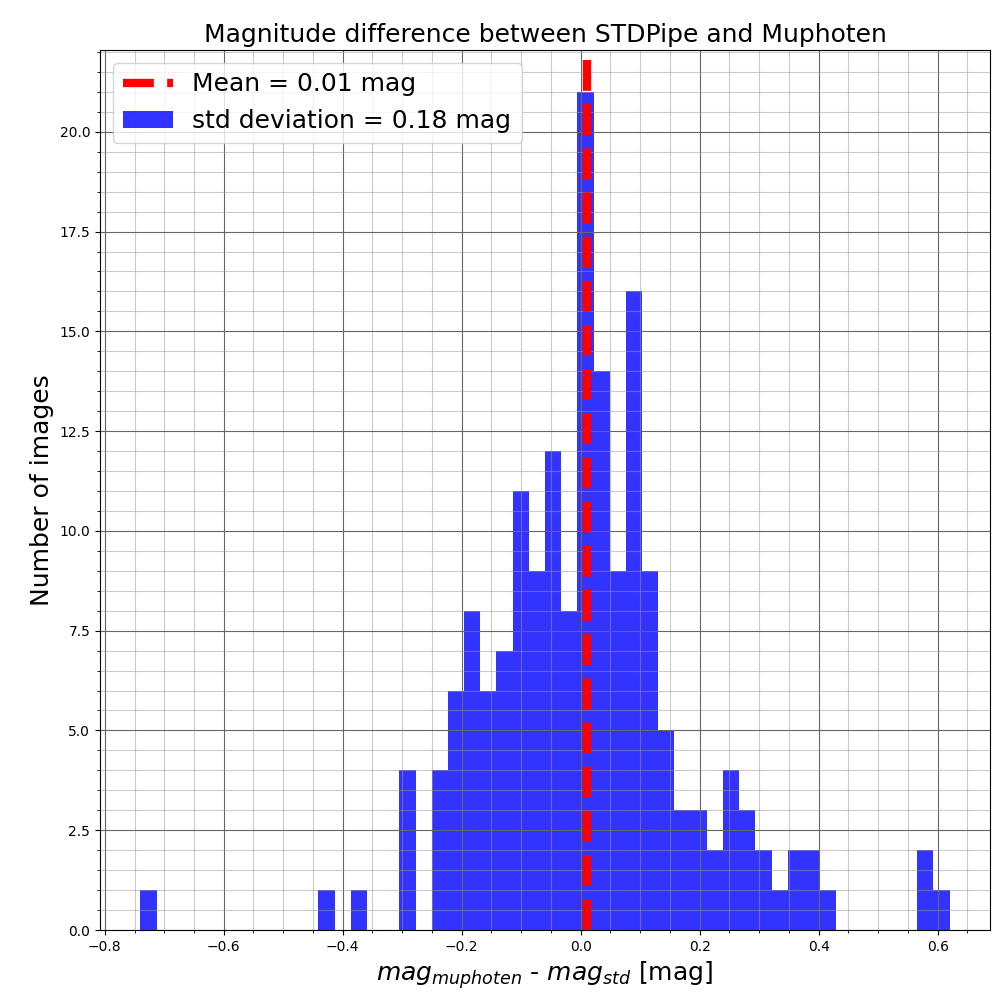}
\caption{Distribution for difference in magnitudes for all the images where the transient is detected, but the unfiltered ones.}
\label{fig:bias-mu-std}
\end{center}
\end{figure}

We processed the images using both the {\sc STDPipe} and {\sc MUphoten} pipelines, thus acquiring two independent sets of measurements based on different photometric models -- the one with a color term for the former, and without it for the latter. While the former is more accurate in theory, it requires the knowledge of a true transient color at the time of measurement in order to convert the result to standard photometric system. As we, in general, do not always know it, for the sake of current analysis we decided to ignore the contribution from color terms and assume the color (either $g^\prime-r^\prime$ or $B-V$, whatever has been used for the photometric calibration of individual frames) to be the zero. 
This introduces a systematic color-dependent error for the two methods to the results of photometry. In order to assess the error, as well as other possible sources of bias, we compare the results of the two different methods in Figure~\ref{fig:bias-mu-std} for all the images where the transient is detected, except for the clear ones and ones using a luminance filter. The latter correspond to a band covering the optical domain from UV to NIR, and is denoted $L$ band in the next sections. The mean difference between {\sc STDPipe} and {\sc MUphoten} magnitudes for these images is less than the 0.1\,mag, which is the typical uncertainty of the  measurements. 
Thus, on average the difference between the results of the two pipelines is compatible with zero and the width of the distribution (0.18 mag) is comparable to the typical accuracy of each measurement, and so we can conclude there are no biases.

\begin{table}
\caption{Summary of the difference mag$_{MU}$ - mag$_{STD}$ used for image reduction, separated by filter.}
\begin{tabular}[t]{*{3}{c}}
\hline
\hline
Band  & Mean difference [mag] & standard deviation [mag] \\ 
\hline
$B$ & 0.2  & 0.2 \\
$V$ & $<$ 0.1  & 0.1 \\
$R_C$ & -0.1 & 0.2 \\
$I_C$ & $<$ 0.1  & 0.2 \\
$g^\prime$    & $<$ 0.1  & 0.1 \\ 
$r^\prime$    & -0.01 & 0.1 \\ 
$i^\prime$    &  0.1 & 0.1 \\
Clear/L & -0.4 & 0.3 \\
\hline                           
\label{tab:band-bias}
\end{tabular}
\end{table}


\subsection{Rapid linear fit and offline classification}
\label{GRANDMAenrichment}

A ``detection'' corresponds to the first \textbf{public} photometric detection of ZTF (defined as {$T_0$}).
Images are grouped with 0.1 day precision. A group contains at least one image.

\textbf{Linear fit method} -- Our goal was to identify transients undergoing rapid decrease or increase in brightness. We applied a linear regression procedure using a maximum likelihood estimation approach to estimate the slopes between two different time 
bins, $t_i$ = T$_0$ +$n_i$ days, where $n_i = 0.1\times i$, and $i$ is the index of a given time bin (see Appendix~\ref{appendix_linearfit})). 
The slopes of the best fit, $a_{r} (t=n)$, are
then computed using data taken with $r$ or $R$ filters in a given time bin. We actually computed three slopes:
\begin{enumerate}
    \item $a_{\textrm{STD},r} (t=n)$, the temporal slope of the transient light curve between the first detection by ZTF at $T_0$ and the time at which GRANDMA detected it by using STDPipe.
    \item $a_{\textrm{MU},r} (t=n)$, the temporal slope of the transient light curve between the first detection by ZTF at $T_0$ and the time at which GRANDMA detected it by using MUphoten.
    \item $a_{\textrm{ZTF},r} (t=n)$, the temporal slope of the transient light curve between the first detection by ZTF at $T_0$ and the next detections by ZTF.
\end{enumerate}

\textbf{Offline source modeling} - to constrain the nature of rapidly evolving optical transients using four light curve models a posteriori. For each of these models, we evaluate its degree of correlation with the observational data. 
We will use a KN model taken from \cite{2017} (Ka2017), a GRB afterglow (TrPi2018, \citealt{TrPi2018}), the nugent-hyper model which creates supernovae light curve (nugent-hyper, \citealt{LeNu2005}) and a shock cooling supernova light curve model (Piro2021, \citealt{PiHa2021}). The ideal model is one that has a regression consistent with the light curve points as assessed by the Bayes factor of the model.

\subsection{Results}
\label{results}

\begin{figure*}
\begin{center}
\includegraphics[width=1.0\textwidth]{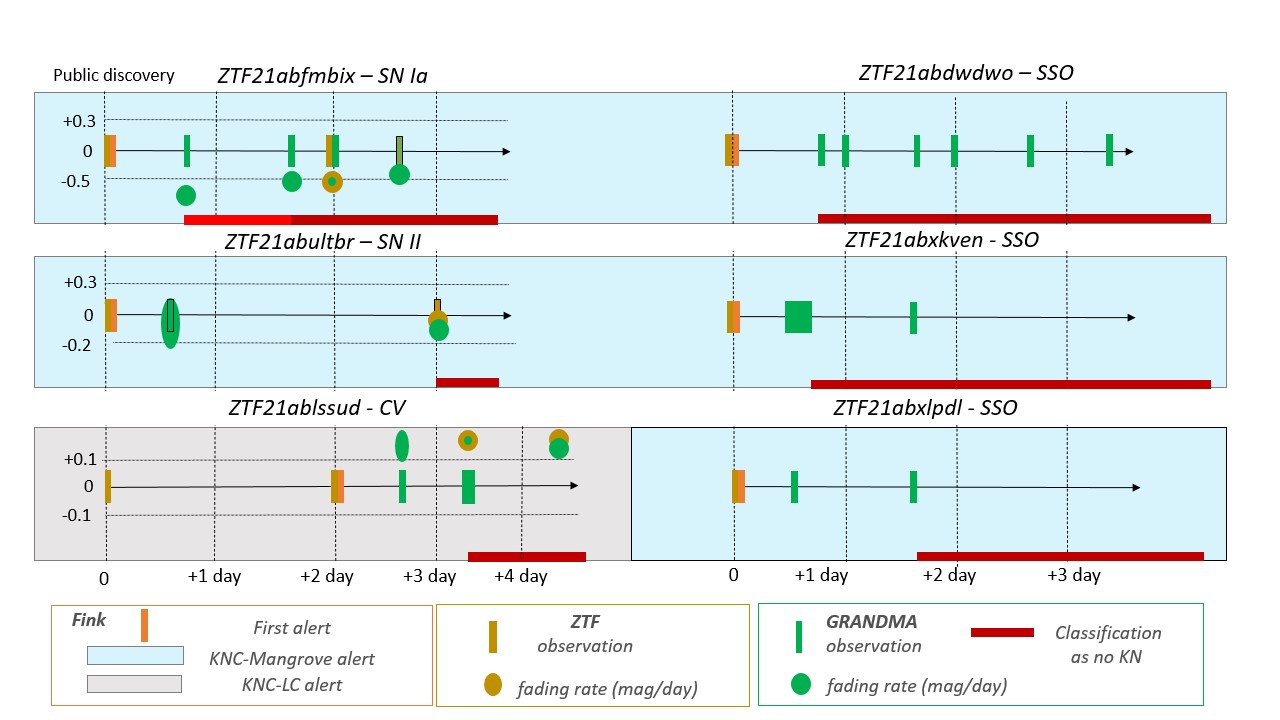}
\caption{Overview of the GRANDMA observations of the six \ztfink\ alerts followed up by professional and amateurs astronomers. The \ztfink\ alerts came from two selection filters of the initial ZTF flux measurements named ``KN-LC'' and ``KN-Mangrove'' (see section~\ref{finkKn}). Displayed in orange are the time-stamps for the first alert provided by \fink\ Gold vertical bars represent the time-stamps of the release of ZTF public data, and green vertical bars represent the GRANDMA data analyzed by STDpipe and MUphoten. The circles present our fading slope estimation using $r'$/$R$ filters (see section~\ref{GRANDMAenrichment}); in gold using only ZTF public data, and in green using ZTF+GRANDMA data. Horizontal red bars show the period when the alert is considered as of no further interest for KN searches. SSO corresponds to Solar System Object, CV to cataclysmic variable, and SN to supernova given by our post-observation analysis months after (see figure~\ref{ztfclass}). We see the potential of the amateur community to distinguish astrophysical events into three categories: moving objects, fast transients (KNe,
GRBs) and slow transients (supernovae, CVs).}
\label{fig:summaryfig}
\end{center}
\end{figure*}

In the following section, we summarize our observations and analyze the results produced by both STDPipe and MUphoten to extract information about the nature of the transients (see Figure~\ref{fig:summaryfig}). We use MUphoten for evaluating the upper limit of the image when no source has been detected by both pipelines independently. If the two pipelines provided inconsistent results for an image that had not been previously rejected, we excluded it from further analysis. Some GRANDMA teams performed their own measurements, but in order to keep a consistent analysis, they are not presented in this article. All our results are presented in Table~\ref{tab:KN_observations}.

\subsubsection{KN-Mangrove alert candidates}

 \href{https://fink-portal.org/ZTF21abdwdwo}{ZTF21abdwdwo} was observed by eleven amateur telescopes with a total of 42 images taken from 0.7 days to 9 days after the public detection (021-06-04 04:27:26 UTC) (see Table~\ref{tab:KN_upper}). The associated alert was sent by \fink\ without any further delay. All images resulted in non-detections with a median upper limit of $17.9\pm0.8$ mag in the $r^\prime$ and $R$-band filters. An unfiltered image taken by the T40-A77DAU telescope 0.8\,days post-detection yielded a 20.7\,mag limit. T-CAT obtained the deepest upper limit ($\sim 21$ mag in $B$- and $G$-band) 3.7 days post-detection (2021-06-07 21:41:04). At a much later date, the VIRT telescope confirmed the previous non-detections on 2021-08-03, 60 days post-detection with an upper detection limit of $R=17.5$\,mag. Our refined analysis showed there was a misassociation of the new source and the possible galaxy (see section~\ref{sec:kn-mangrove}). In summary, GRANDMA follow-up classified the new source as a Solar System Object 0.8 day after ZTF discovery.

\href{https://fink-portal.org/ZTF21abfmbix}{ZTF21abfmbix} (later renamed SN 2021pkz, associated with the galaxy 2MASS-12551554+0253477 at 38 Mpc) was observed by 13 different amateur telescopes and FRAM-Auger with a total of 29 images taken between 0.7 and 50.8 days after T$_0$ (2021-06-11 05:14:49). The associated alert was sent by \fink\ with a 1.5 hour delay. Two days before, ZTF did not detect any source brighter than $g\prime>20.5$. A positive detection was found in 24 of the 29 images, using both the STDPipe and MUphoten pipelines. These images were taken in six different filters and also with no filter from 0.7 days to 29.7 days post-detection. The next public ZTF measurement ($g^\prime=17.1$ mag, $r^\prime=16.9$ mag) was delivered two days after the first detection. We note that 17 of 25 photometric measurements (68\%) delivered from STDPipe and MUphoten are consistent within $0.1\sigma$. We also note a maximum deviation of 0.6 mag in the $L$ band and for images taken with no filter, when comparing the results of both pipelines. According to the results obtained with both pipelines, the data clearly show a transient in rising phase in the $r^\prime$ band. We evaluated the slopes of the luminosity rate (see Section~\ref{GRANDMAenrichment}) as: $a_{\textrm{STD},r} (t=0.7) = -0.7 \pm 0.2$ mag/day and $a_{\textrm{MU},r} (t=0.7) = -0.9 \pm 0.2$ mag/day. While the luminosity rate estimations differ between STDPipe and MUphoten, it is clear that the source is brightening. We also found agreement between the slope of the luminosity rate using the different pipelines for $\textnormal{t}=1.7$ ($-0.5\pm0.1$ and $-0.7\pm0.1$ mag/day for STD and MU, respectively), and the value obtained from the public data at $\textnormal{t}=2.0$ days ($a_{\textrm{ZTF},r} = -0.5 \pm 0.1$ mag/day). This example shows the benefit of advanced measurements (+16~h first estimation, and 40~h confirmation) during the rising phase of the transient, and this will be useful for any decision on spectroscopy observations before night-time in the Americas. This depends on our capacity to promptly analyze the data as soon as it is acquired. A two-day long rise in the $r^\prime$ band is atypical for standard KNe at 38 Mpc and could only be related to very particular ejecta configurations during a NS-BH collision \citep{2019MNRAS.489.5037B}. Also, we note an independent spectroscopic measurement by the ZTF Spectral Energy Distribution Machine (SEDM) on T$_0+0.1$ days, classifying the source as a \href{https://www.wis-tns.org/object/2021pkz}{supernova Ia}.

\href{https://fink-portal.org/ZTF21abultbr}{ZTF21abultbr} was observed by three amateur telescopes and by the Abastumani-T70, providing a total of eight images from 0.6 day to 2.5 days after T$_0$ (2021-08-21 02:44:39.100 UTC). The associated alert was sent by \fink\ with a 10 minute latency with respect to the public detection. The source was presumably associated with \href{http://leda.univ-lyon1.fr/ledacat.cgi?o=UGC04104}{HyperLEDA/UGC04104}, located at 89 Mpc. 
Three of the eight images, taken with two different telescopes, led to a positive detection of the object using the STDPipe and MUphoten pipelines.
These images were taken with in $R$ and $G$ bands, and also with unfiltered observations. The next public ZTF measurement ($r^\prime=18.7$\,mag) was delivered three days after the first detection of the source. We note the consistency of photometric measurements between STDPipe and MUphoten, except for images taken with no filter. At $\textnormal{t}=0.6$ day, the luminosity rate (see Sect.~\ref{GRANDMAenrichment}) exhibits the following slopes: $a_{\textrm{STD},r} (t=0.6) = -0.2 \pm 0.4$ mag/day and $a_{\textrm{MU},r} (t=0.6) = -0.0 \pm 0.4$ mag/day. The value obtained from the public data at $\textnormal{t}=3.0$ days ($a_{\textrm{ZTF},r^\prime} (t=3.0) = -0.0 \pm 0.1$ mag/day). These measurements were insufficient to conclude the nature of the transient and how it evolved from 0.6 to 3.0 days.
The monitoring of the source by ZTF was interrupted between $\textnormal{t}=3.0$ to $\textnormal{t}=21.0$ days. The source was independently classified as a \href{https://www.wis-tns.org/object/2021wqd}{supernova II} after 21 days by the ZTF SEDM. 

\href{https://fink-portal.org/ZTF21abxkven}{ZTF21abxkven} was observed simultaneously by the Abastunami/T70, TRT-SRO, Lisnyky/Schmidt-Cassegrain, and FRAM-CTA telescopes, as well as nine different amateur telescopes with a total of 17 images taken between 0.5 and 11.7 days after T$_0$ (2021-09-03 08:28:07). It was presumably associated with \href{http://leda.univ-lyon1.fr/ledacat.cgi?o=UGC12816}{HyperLEDA- UGC12816} located at 80 Mpc based on the KN-Mangrove filter. The associated alert was sent by \fink\ without any further delay. With no more alert data sent for this location on the sky, \fink\ classified this transient independently as a Solar System object \citep{10.1093/mnras/staa3602}. All images resulted in non-detections with a median upper limit of 20.6 mag in underfiltered images within the first day of observation. A clear image (no filter) was taken at 0.7 day after T$_0$ by T-STSOPHIE, with a 20.6 mag upper limit. In summary, the GRANDMA follow-up ruled out any possible existing KN within 80 Mpc at T0+0.7 day.

\href{https://fink-portal.org/ZTF21abxlpdl}{ZTF21abxlpdl} was observed simultaneously by the Abastunami/T70, FRAM-CTA, and TRT-SRO telescopes, as well as seven amateur telescopes. A total of 15 images were taken from 0.5 to 13.9 days after the first public detection (2021-09-03 08:59:15). The alert was sent the same day as ZTF21abxkven, so the participation was limited for both targets. The transient was presumably associated with NGC 105, located at 79 Mpc based on our KN-Mangrove filter. The associated alert was sent by \fink\ without any further delay. All images resulted in non-detections with a median upper limit of $20.0$ in unfiltered observations (see some examples in Table~\ref{tab:KN_upper}). A clear image was taken 1.7 days after the alert using the T40-A77DAU, with an upper limit estimated at $>20.7$\,mag. In summary, the GRANDMA follow-up ruled out any KN within 80 Mpc at T0+1.7 day with luminosity decay rate $<1$ mag/day in the $r^\prime$ band. Some scenarios involving both BNS and NSBH collisions can produce steeper decay rates $>1$ mag/day in the $r^\prime$ band, although these cases are extremely rare (see \citealt{2019MNRAS.489.5037B}).

\subsubsection{KN-LC alert candidates}

\begin{figure}
\begin{center}
\includegraphics[width=0.5\textwidth]{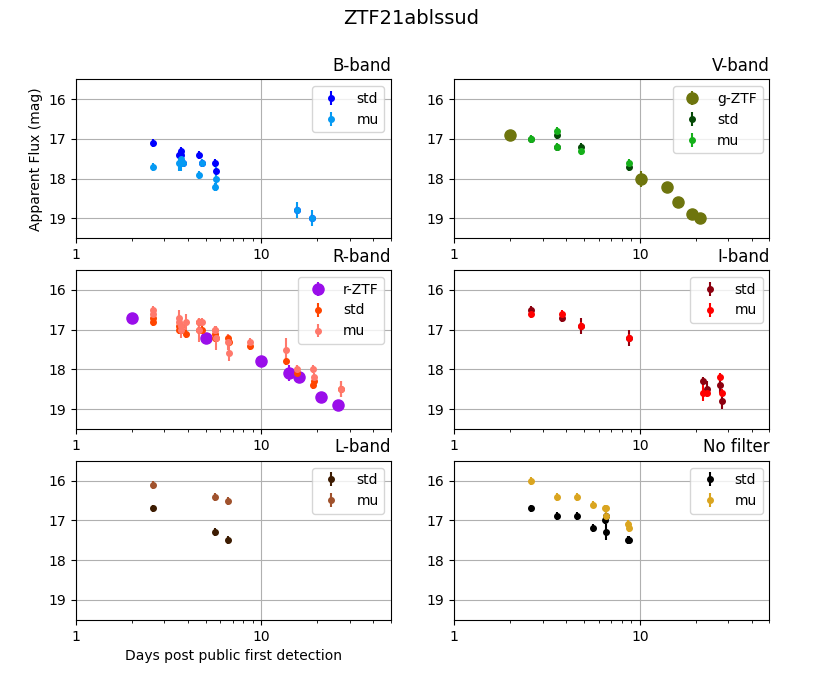}
\caption{ZTF21ablssud light curves with ZTF as well as the GRANDMA observations from 17 different telescopes. The data were mostly taken by amateur astronomers before $t=T_0+10$ days, and by professionals for $t>T_0+10$ days. STDPipe and MUphoten measurements are in agreements for $V$, $R_C$, and $I_C$ bands. In the $L$ filter as well as in unfiltered images, MUphoten adds flux from $g^\prime$- and $r^\prime$-band Pan-STARRS images while STDPipe treats them as $R_C$-band images. We show that the GRANDMA measurements are consistent with the ZTF ones, allowing for filling in the light curve gaps. However the use of non-standard filters by some amateur astronomers (especially the $B$ band of the T-CAT instrument) can lead to discrepancies between measurements up to 0.5 mag. 
}
\label{fig:ZTF21ablssudLC}
\end{center}
\end{figure}

\href{https://fink-portal.org/ZTF21ablssud}{ZTF21ablssud} was observed simultaneously by the TRT-SRO, Lisnyky/Schmidt-Cassegrain, Tibet-50, and UBAI-NT60 telescopes, as well as 17 different amateur telescopes, leading to a total of 141 images taken between 2.6 to 26.7 days after T$_0$ (2021-07-16 21:11:45). The associated alert from the KN-LC filter was sent by \fink\, two days after the discovery with a probability of 53\% classification (see section~\ref{sec:kn-lc}). Other scores were delivered by \fink\: Early SN (5\%), Supernova SN Ia vs non-IA SN (73\%) and  SN Ia and Core-Collapse vs non-SN (39\%). Observations started 0.6 days after the \fink\ alert. A total of 66 images taken with eight different filters (and additional unfiltered images) from 17 telescopes exhibited a positive detection of the object using both the STDPipe and MUphoten pipelines. The next public ZTF measurement post-\fink\ alert ($r^\prime=17.2$\,mag in the $r^\prime$ band) was delivered five days after the first detection. 
For $\textnormal{t}=3.7$ days, we derived the slope of the luminosity decay rate (section~\ref{GRANDMAenrichment}) as $0.15\pm0.1$ mag/day for both STDPipe and MUphoten. The slopes obtained from the ZTF public data are $a_{\textrm{ZTF},r} = 0.15 \pm 0.1$ mag/day at $\textnormal{t}=2.0$ day and $\textnormal{t}=5.0$ day.
We ruled out the association with a standard KN resulting from the coalescence of two binary neutron stars. No spectra have been reported in the literature. \fink\ scored the transient as Supernova (79\%) after 30 days of observations by ZTF. GRANDMA accumulated an important collection of data for  $\textnormal{t}<10$ days compared to the available data in the literature
(see Figure~\ref{fig:ZTF21ablssudLC}). 
Based on the color evolution and the location of the source close to the galactic plane, we proposed that the source corresponds to a cataclysmic variable event. 
In addition, the light curve fitting of ZTF data ruled out the nature of the source as a KN, GRB afterglow, or supernova event (see section~\ref{ztfclass}).

\subsubsection{Training alert candidates}

Here, we briefly describe our observations on ``training alerts'' distributed by \fink. The alerts are produced via different channels: supernova and KN candidates. They were not only scheduled for Fridays and were proposed for observing on a best-effort basis by the \fink\ team for practice.

\href{https://fink-portal.org/ZTF21abfaohe}{ZTF21abfaohe/SN~2021pfs} and  \href{https://fink-portal.org/ZTF21abbzjeq}{ZTF21abbzjeq/ SN~2021mwb} were classified as supernovae (see TDS for external reference and section~\ref{ztfclass}). \href{https://fink-portal.org/ZTF21abotose}{ZTF21abotose/SN~2021ugl} is a supernova IIb. The three sources distributed via \fink\ were selected to test the data reduction capabilities of GRANDMA. Photometric results are reported in Appendix~\ref{obsApp}.

\href{https://fink-portal.org/ZTF21abyplur}{ZTF21abyplur} was only observed by the Tibet-50 with four images beginning 9.5 days after the first public detection reported on 2021-09-07 09:12:26. The associated alert was sent by \fink\ without any further delay but the amateur astronomers were not notified because it was outside of our Friday exercise. The alert-host association (HyperLEDA-PGC1115282) is within 10\arcsec, almost at the limit of a positive association; it might have been incorrectly associated with the galaxy. \fink\ classified this object independently as a Solar System object, probably attached to Solar System object number 22327 \citep{10.1093/mnras/staa3602}. The images taken by the Tibet-50 could not rule out the case of a fast transient with an upper limit of $g^\prime>18.3$ mag, at $\textnormal{t}=\textnormal{T}_0+9.5$ days.

\href{https://fink-portal.org/ZTF21absvlrr}{ZTF21absvlrr} was observed by six amateur telescopes and the Terskol/Zeiss-600, TRT-SBO, TRT-SRO, and Abastumani-T70 telescopes, leading to a total of 24 images taken from 1 to 59 days after T$_0$ (2021-08-12 09:52:43). The associated alert from the KN-Mangrove filter was sent by \fink\ with a 2.3 hour delay, but distributed to GRANDMA as an practical exercise about 0.9 days afterwards.
The source was presumably associated with \href{http://leda.univ-lyon1.fr/ledacat.cgi?o=ESO540-025}{HyperLEDA/ESO540-025} located at 89 Mpc. A total of 19 images taken with seven telescopes confirmed a positive detection of the source using both STDPipe and MUphoten pipelines.
These images were taken in $B$, $V$, $r^\prime$, $R$, as well as unfiltered. The next public ZTF measurement ($g\prime=17.6$ mag, $r\prime=17.7$ mag) was delivered  two days after the first detection. We note the consistency of photometric measurements between STDPipe and MUphoten, except for images taken with no filter. According to the results obtained with both pipelines, the data clearly shows a transient in the rising phase in the $r\prime$ band, based on the slopes of the light curves at $\textnormal{t}=1.0$ day ($-0.1\pm0.2$ mag/day for both STD and MU), and at $\textnormal{t}=1.7$ days ($-0.5\pm0.1$ and $-0.7\pm0.1$ mag/day for STD and MU, respectively). At $\textnormal{t}=2.0$ days, the slope obtained from the ZTF public data is $a_\textrm{ZTF,r}=-0.5\pm0.1$ mag/day. In addition, we also observed a two-day long rise of the source in $B$ ($-0.7\pm0.2$ and $-0.8\pm0.2$ mag/day for STD and MU, respectively, at $\textnormal{t}=2.0$ days) and in the $g\prime$-band ($a_\textrm{ZTF;g}(t = 2.0) = -0.6\pm0.1$ mag/day). The source, at an early stage, did not behave in a manner similar to AT2017gfo. We note an independent spectroscopic measurement by the ZTF SEDM on T$_0+1$ day, classifying the source as a \href{https://www.wis-tns.org/object/2021vtq}{supernova Ia}.

 \href{https://fink-portal.org/ZTF21acceboj}{ZTF21acceboj/SN 2021yyg} was only observed by Tibet-50 and Terskol-600 telescopes with a total of 11 images taken after 1.4 days from the first public detection reported on 2021-09-14 11:04:25. The associated alert from the KN-Mangrove filter was sent by \fink\ without any further delay but the amateur astronomers were not notified since it was out of our Friday triggering schedule. From linear regression fitting (see section~\ref{GRANDMAenrichment}) we obtained $a_{\textrm{STD},r} (t=1.4) = -0.3 \pm 0.1$ mag/day, and $a_{\textrm{MU},r} (t=1.4) = -0.4 \pm 0.1$ mag/day. Using public data we measured $a_{\textrm{ZTF},r} (t=2.0) = -0.3 \pm 0.1$ mag/day. The source clearly exhibits a two-day rise in the $r^\prime$ band that ruled out
ZTF21acceboj as a KN stemming from the coalescence of two binary neutron stars. The Global SN Project (with LCO) classified the source as a \href{https://www.wis-tns.org/object/2021yyg}{supernova IIp} one day after than the first public detection.
 
\subsubsection{Constraining the nature of transients with offline results}
\label{ztfclass}
The objective of the ``ReadyforO4'' campaign was to gather information on the nature of transient events and their evolution. Events were followed during weekends, but not monitored over several weeks. Hence our GRANDMA sample covered essentially two epochs: when the alert was received by our amateur astronomers, and when the professional telescopes joined the campaign in July. In this sense, we trained our team on the ZTF data available to confirm the nature of the sources we followed up with GRANDMA using models described in section~\ref{GRANDMAenrichment}. The ZTF data are used to validate the models by generating the logarithm of the Bayes factor, the level of correlation of the model fit with the data, and increase our knowledge of the physical processes following the observation.

In Table~\ref{nmmaTable}, we present the results obtained for each transient clearly identified as a non-moving object through the analysis of the light curve from ZTF data.
Using the models described above, we identify ZTF21abfmbix, ZTF21absvlrr, ZTF21abultbr, ZTF21abfaohe, ZTF21abfmbix, ZTF21abbzjeq and ZTF21acceboj as typical supernova candidates.
ZTF21abotose is both consistent with shock cooling and GRB afterglow models at early times. However, the increase of its brightness at later times is in good agreement with the shock cooling scenario only.
ZTF21ablssud is well-fit by a GRB model, however, due to its galactic latitude ($\ell=5.7$ deg) and similarity to many other examples of cataclysmic variables in the literature, it is likely a cataclysmic variable. 




\begin{table*}
\begin{center}
\caption{Results of the simulations of transients with rapid evolution, in order to validate or reject the concordance of each transient using ZTF data with the four models:  KNe (Ka2017), supernova (nugent-hyper), GRB afterglows (TrPi2018) and shock cooling (Piro2021). Note that the Bayes factors are evaluated logarithmically.}
\begin{tabular}[t]{*{8}{c}}
\hline
\hline
Transients  & Ka2017  & TrPi2018  & nugent-hyper & Piro2021 & light curve  \\ 
 
\hline

{ZTF21abfmbix} &$-12.38$   & $ -15.95$    & $-9.18$   & $-10.1$       & {Supernova Ia}\\

{ZTF21absvlrr}  &  $-9.78$  & $-16.73$   & $ -9.91$    &  $-11.07$ &  {Supernova Ia} \\

{ZTF21abultbr}      & $-2.73$ & -$9.14$  &$ -5.24$   & $-4.76$    & {Supernova II} \\

{ZTF21ablssud}       &$ -6.32$   & $-11.58 $  &$-9.83$  &$ -9.41$ &{Cataclysmic Variable} \\

{ZTF21abfaohe}    &$-12.3$  &$-10.98$   & $-7.47$ & $8.67$     &  {Supernova Ia}     \\

{ZTF21abbzjeq}   &$-8.22$   & $-11.41 $    & $-7.49$  &$ -8.47$ & {Supernova Ia }  \\
 
{ZTF21acceboj}     & $-16.52$  & $-19.44$    &$-14.52$ & $-15.6$ & {Supernova IIb}\\
                                          
{ZTF21abotose}    &$ -6.37$  &$-10.62$   & $-7.41$   & $-7.49$ & {Shock Cooling - Supernova IIp}      \\
          
\hline                           

\end{tabular}
\label{nmmaTable}
\end{center}

\end{table*}

\section{Conclusions}
\label{conclusions}

In this paper we describe GRANDMA's ``ReadyforO4'' campaign to search for serendipitous KNe using ZTF public data filtered by \fink\ and followed-up with GRANDMA facilities. Eight objects in total were observed, using 26 amateur telescopes and eleven professional telescopes (34 of which provided data analyzed in this work). Through \fink\ and follow-up, no KNe were identified, instead we classified four of these objects as asteroids, while the remaining were classified as cataclysmic variables and supernovae. In addition to demonstrating a number of challenges for observing transient sources with a variety of telescopes of different apertures, filters, and configurations, we achieved a number of successes.

\begin{itemize}
\item \textbf{Reaction time} --  We obtained the first images from the amateur community less than 16 hours after the \fink\ alert. This delay is due to the fact that most of our amateur community is located in Europe (especially in France). In the future we hope to connect with other amateurs across the world, although language barriers remain a challenge. 
\item \textbf{Data acquisition and filters} -- The GRANDMA facilities are mostly equipped with red filters, which are excellent for the characterization of KNe events. Our campaign showed that amateur astronomers reached a depth of 21 mag with a variety of filter configurations. These astronomers will be an asset during O4, for which we might expect AT2017gfo-type KN events peaking at $\sim21$ mag for events located at the observational horizon of the GW interferometers \citep{2022ApJ...924...54P}. However, the use of filter sets (Johnson-Cousins, $L$-, $w$- and $o$-filters) different from those of the SDSS, which was our reference catalog for this analysis, remains problematic for the characterization of the sources. For this reason, we found that the use of two standard data reduction pipelines (STDPipe and MUPHOTEN) is a way to standardize results (as compared to allowing the individual telescope teams to reduce their own data). We observed a difference in magnitude of less than $0.2$ mag for Johnson-Cousins and SDSS filters. Based on this experience, we would like to motivate our community to use only SDSS filters to have a homogeneous set of data. We also noted that images taken with no filter are particularly useful for guiding the observers for their data acquisition. We finally experienced miscommunication on products distributed to GRANDMA, as some observers provided pre-stacked images while others were submitted as individual images. This greatly reduced our ability to analyze the images at low latency, and so correcting this will be a focus for future campaigns.
\item \textbf{Quality of the data sample} -- Among the eight transients, a total of 450 images were taken by GRANDMA partners. However, we detected a source and were able to perform astrometry and photometry on only 180 images (40\%). To improve this efficiency, we need amateur astronomers to gain expertise in their data acquisition and validation before distribution to the network. For example, some data had artifacts and ``star'' tracks, blur, or galaxy saturation that could easily be identified with more experience. Another challenge was the dispersion of format and keywords employed in the files, which slowed down our analysis. This will be partially solved in future runs with a set of required keywords, including the time of the start of the observation, the name of the telescope, the filter used and the filter system.    
\item \textbf{Classification} -- We demonstrated the potential of the amateur community to distinguish astrophysical events into three categories: moving objects, fast transients (KNe, GRBs) and slow transients (supernovae, CVs). We based our rapid fast transient classification on the decay rate of the optical sources, which should reach a maximum of at least 0.3 mag/day for KNe. The images posted by the professional and amateurs observers within the first 2 days after the \fink\ alert helped to categorize the candidates, which were discovered at an average magnitude of $r^\prime18.1\pm1.0$ mag. This classification improved on the timing relative to routine observations by ZTF during the American night. In addition, we also developed rapid and sophisticated modeling tools that can be applied to GRANDMA data during the O4 observing run. If we are able to run data reduction in near real-time, we would be able to filter the most probable candidates a few hours before the second night of observation in the Americas after a GW alerts. This will allow us to trigger spectroscopic observations with higher confidence during the O4 observing run, mitigating the use of valuable resources on non-viable candidates.
\end{itemize}

Overall, we consider the first ``ReadyforO4'' campaign to be a success which can be built upon to allow the amateur community to partake in cutting-edge astrophysical science once the fourth observing run of the LIGO-Virgo-KAGRA detectors begins in late 2022/early 2023.




\section*{Data availability}
The data underlying this article will be shared on reasonable request to the corresponding author.

\section{Acknowledgements}
SA and CL acknowledge the financial support of the Programme National Hautes Energies (PNHE). SA acknowledges the financial support of CNES. SA is grateful for financial support from the Nederlandse Organisatie voor Wetenschappelijk Onderzoek (NWO) through the VIDI (PI: Nissanke). SA dedicates her contribution to Rayan Ouram, who is a source of inspiration for bravity and humanity for GRANDMA. DT acknowledges the financial support of CNES post-doctoral program. UBAI acknowledges support from the Ministry of Innovative Development through projects FA-Atech-2018-392 and VA-FA-F-2-010. RI acknowledges Shota Rustaveli National Science Foundation (SRNSF) grant No - RF/18-1193. TAROT has been built with the support of the Institut National des Sciences de l'Univers, CNRS, France. 
MP, SK and MM are supported by European Structural and Investment Fund and the Czech Ministry of Education, Youth and Sports (Projects CZ.02.1.01/0.0/0.0/16\_013/0001403, CZ.02.1.01/0.0/0.0/18\_046/0016007 and CZ.02.1.01/0.0/0.0/15\_003/0000437). The FRAM telescope is also supported by the Czech Ministry of Education, Youth and Sports (projects LM2015046, LM2018105, LTT17006).
NBO and DM acknowledge financial support from NASA MUREP MIRO award 80NSSC21M0001, NASA EPSCoR award 80NSSC19M0060, and NSF EiR award 1901296. PG acknowledges financial support from NSF EiR award 1901296. 
DAK acknowledges support from Spanish National Research Project RTI2018-098104-J-I00 (GRBPhot)
XW is supported by the National Science Foundation of China (NSFC grants 12033003 and 11633002), the Scholar Program of Beijing Academy of Science and Technology (DZ:BS202002), and the Tencent Xplorer Prize. 
The GRANDMA consortium thank the amateur participants to the \textit{kilonova-catcher} program. The \textit{kilonova-catcher} program is supported by the IdEx Universit\'e de Paris, ANR-18-IDEX-0001. This research made use of the cross-match service provided by CDS, Strasbourg. MC acknowledges support from the National Science Foundation with grant numbers PHY-2010970 and OAC-2117997. GR acknowledges financial support from the Nederlandse Organisatie voor Wetenschappelijk Onderzoek (NWO) through the Projectruimte and VIDI grants (PI: Nissanke). Thanks to the National Astronomical Research Institute of Thailand (Public Organization), based on observations made with the Thai Robotic Telescope under program ID TRTC08D\_005 and TRTC09A\_002.
S. Leonini thanks M. Conti, P. Rosi, and L. M. Tinjaca Ramirez.
SA thanks Etienne Bertrand and le ``Club des C\'eph\'eides'' for their observations of ZTF21abxkven.

\bibliographystyle{mnras}
\bibliography{references}

\begin{thebibliography}{}
\makeatletter
\relax
\def\mn@urlcharsother{\let\do\@makeother \do\$\do\&\do\#\do\^\do\_\do\%\do\~}
\def\mn@doi{\begingroup\mn@urlcharsother \@ifnextchar [ {\mn@doi@}
  {\mn@doi@[]}}
\def\mn@doi@[#1]#2{\def\@tempa{#1}\ifx\@tempa\@empty \href
  {http://dx.doi.org/#2} {doi:#2}\else \href {http://dx.doi.org/#2} {#1}\fi
  \endgroup}
\def\mn@eprint#1#2{\mn@eprint@#1:#2::\@nil}
\def\mn@eprint@arXiv#1{\href {http://arxiv.org/abs/#1} {{\tt arXiv:#1}}}
\def\mn@eprint@dblp#1{\href {http://dblp.uni-trier.de/rec/bibtex/#1.xml}
  {dblp:#1}}
\def\mn@eprint@#1:#2:#3:#4\@nil{\def\@tempa {#1}\def\@tempb {#2}\def\@tempc
  {#3}\ifx \@tempc \@empty \let \@tempc \@tempb \let \@tempb \@tempa \fi \ifx
  \@tempb \@empty \def\@tempb {arXiv}\fi \@ifundefined
  {mn@eprint@\@tempb}{\@tempb:\@tempc}{\expandafter \expandafter \csname
  mn@eprint@\@tempb\endcsname \expandafter{\@tempc}}}

\bibitem[\protect\citeauthoryear{{Abbott} et~al.,}{{Abbott}
  et~al.}{2017a}]{LSC_BNS_2017PhRvL}
{Abbott} B.~P.,  et~al., 2017a, \mn@doi [\prl]
  {10.1103/PhysRevLett.119.161101}, \href
  {https://ui.adsabs.harvard.edu/abs/2017PhRvL.119p1101A} {119, 161101}

\bibitem[\protect\citeauthoryear{{Abbott} et~al.,}{{Abbott}
  et~al.}{2017b}]{LSC_MM_2017ApJ}
{Abbott} B.~P.,  et~al., 2017b, \mn@doi [\apjl] {10.3847/2041-8213/aa91c9},
  \href {https://ui.adsabs.harvard.edu/abs/2017ApJ...848L..12A} {848, L12}

\bibitem[\protect\citeauthoryear{{Alexander} et~al.,}{{Alexander}
  et~al.}{2017}]{2017ApJ...848L..21A}
{Alexander} K.~D.,  et~al., 2017, \mn@doi [\apjl] {10.3847/2041-8213/aa905d},
  \href {https://ui.adsabs.harvard.edu/abs/2017ApJ...848L..21A} {848, L21}

\bibitem[\protect\citeauthoryear{{Almualla}, {Coughlin}, {Anand}, {Alqassimi},
  {Guessoum}  \& {Singer}}{{Almualla} et~al.}{2020}]{2020MNRAS.495.4366A}
{Almualla} M.,  {Coughlin} M.~W.,  {Anand} S.,  {Alqassimi} K.,  {Guessoum} N.,
    {Singer} L.~P.,  2020, \mn@doi [\mnras] {10.1093/mnras/staa1498}, \href
  {https://ui.adsabs.harvard.edu/abs/2020MNRAS.495.4366A} {495, 4366}

\bibitem[\protect\citeauthoryear{Almualla et~al.,}{Almualla
  et~al.}{2021}]{Almualla_2021}
Almualla M.,  et~al., 2021, \mn@doi [Monthly Notices of the Royal Astronomical
  Society] {10.1093/mnras/stab1090}, 504, 2822–2831

\bibitem[\protect\citeauthoryear{{Andreoni}, {Ackley}, {Cooke}
  et~al.}{{Andreoni} et~al.}{2017}]{Andreoni_2017PASA}
{Andreoni} I.,  {Ackley} K.,  {Cooke} J.,   et~al., 2017, \mn@doi [PASA]
  {10.1017/pasa.2017.65}, \href
  {https://ui.adsabs.harvard.edu/\#abs/2017PASA...34...69A} {34, e069}

\bibitem[\protect\citeauthoryear{{Andreoni} et~al.,}{{Andreoni}
  et~al.}{2019}]{2019PASP..131f8004A}
{Andreoni} I.,  et~al., 2019, \mn@doi [\pasp] {10.1088/1538-3873/ab1531}, \href
  {https://ui.adsabs.harvard.edu/abs/2019PASP..131f8004A} {131, 068004}

\bibitem[\protect\citeauthoryear{{Andreoni} et~al.,}{{Andreoni}
  et~al.}{2021a}]{2021arXiv210606820A}
{Andreoni} I.,  et~al., 2021a, arXiv e-prints, \href
  {https://ui.adsabs.harvard.edu/abs/2021arXiv210606820A} {p. arXiv:2106.06820}

\bibitem[\protect\citeauthoryear{{Andreoni} et~al.,}{{Andreoni}
  et~al.}{2021b}]{2021ApJ...918...63A}
{Andreoni} I.,  et~al., 2021b, \mn@doi [\apj] {10.3847/1538-4357/ac0bc7}, \href
  {https://ui.adsabs.harvard.edu/abs/2021ApJ...918...63A} {918, 63}

\bibitem[\protect\citeauthoryear{{Antier} et~al.,}{{Antier}
  et~al.}{2020a}]{GRANDMAO3A}
{Antier} S.,  et~al., 2020a, \mn@doi [\mnras] {10.1093/mnras/stz3142}, \href
  {https://ui.adsabs.harvard.edu/abs/2020MNRAS.492.3904A} {492, 3904}

\bibitem[\protect\citeauthoryear{{Antier} et~al.,}{{Antier}
  et~al.}{2020b}]{GRANDMA03B}
{Antier} S.,  et~al., 2020b, \mn@doi [\mnras] {10.1093/mnras/staa1846}, \href
  {https://ui.adsabs.harvard.edu/abs/2020MNRAS.497.5518A} {497, 5518}

\bibitem[\protect\citeauthoryear{{Arcavi} et~al.,}{{Arcavi}
  et~al.}{2017}]{2017Natur.551..210A}
{Arcavi} I.,  et~al., 2017, \mn@doi [\nat] {10.1038/nature24030}, \href
  {https://ui.adsabs.harvard.edu/abs/2017Natur.551..210A} {551, 210}

\bibitem[\protect\citeauthoryear{Barbary}{Barbary}{2016}]{sep}
Barbary K.,  2016, \mn@doi [Journal of Open Source Software]
  {10.21105/joss.00058}, 1, 58

\bibitem[\protect\citeauthoryear{{Bauswein}, {Just}, {Janka}  \&
  {Stergioulas}}{{Bauswein} et~al.}{2017}]{2017ApJ...850L..34B}
{Bauswein} A.,  {Just} O.,  {Janka} H.-T.,   {Stergioulas} N.,  2017, \mn@doi
  [\apjl] {10.3847/2041-8213/aa9994}, \href
  {https://ui.adsabs.harvard.edu/abs/2017ApJ...850L..34B} {850, L34}

\bibitem[\protect\citeauthoryear{{Becker}}{{Becker}}{2015}]{hotpants}
{Becker} A.,  2015, {HOTPANTS: High Order Transform of PSF ANd Template
  Subtraction} (\mn@eprint {ascl} {1504.004})

\bibitem[\protect\citeauthoryear{{Bellm} et~al.,}{{Bellm}
  et~al.}{2019}]{2019PASP..131f8003B}
{Bellm} E.~C.,  et~al., 2019, \mn@doi [\pasp] {10.1088/1538-3873/ab0c2a}, \href
  {https://ui.adsabs.harvard.edu/abs/2019PASP..131f8003B} {131, 068003}

\bibitem[\protect\citeauthoryear{Berger}{Berger}{2014}]{Berger:2013jza}
Berger E.,  2014, \mn@doi [Ann. Rev. Astron. Astrophys.]
  {10.1146/annurev-astro-081913-035926}, 52, 43

\bibitem[\protect\citeauthoryear{{Berger}, {Fong}  \& {Chornock}}{{Berger}
  et~al.}{2013}]{2013ApJ...774L..23B}
{Berger} E.,  {Fong} W.,   {Chornock} R.,  2013, \mn@doi [\apjl]
  {10.1088/2041-8205/774/2/L23}, \href
  {https://ui.adsabs.harvard.edu/abs/2013ApJ...774L..23B} {774, L23}

\bibitem[\protect\citeauthoryear{{Bertin}}{{Bertin}}{2006}]{scamp}
{Bertin} E.,  2006, {Automatic Astrometric and Photometric Calibration with
  SCAMP}.
p.~112

\bibitem[\protect\citeauthoryear{{Bertin}}{{Bertin}}{2011a}]{2011psfex}
{Bertin} E.,  2011a, in {Evans} I.~N.,  {Accomazzi} A.,  {Mink} D.~J.,   {Rots}
  A.~H.,  eds,  Astronomical Society of the Pacific Conference Series Vol. 442,
  Astronomical Data Analysis Software and Systems XX. p.~435

\bibitem[\protect\citeauthoryear{{Bertin}}{{Bertin}}{2011b}]{psfex}
{Bertin} E.,  2011b, {Automated Morphometry with SExtractor and PSFEx}.
p.~435

\bibitem[\protect\citeauthoryear{{Bertin} \& {Arnouts}}{{Bertin} \&
  {Arnouts}}{1996}]{sextractor}
{Bertin} E.,  {Arnouts} S.,  1996, \mn@doi [\aaps] {10.1051/aas:1996164}, \href
  {https://ui.adsabs.harvard.edu/abs/1996A%26AS..117..393B} {117, 393}

\bibitem[\protect\citeauthoryear{{Bertin}, {Mellier}, {Radovich}, {Missonnier},
  {Didelon}  \& {Morin}}{{Bertin} et~al.}{2002}]{2002ASPC..281..228B}
{Bertin} E.,  {Mellier} Y.,  {Radovich} M.,  {Missonnier} G.,  {Didelon} P.,
  {Morin} B.,  2002, in {Bohlender} D.~A.,  {Durand} D.,   {Handley} T.~H.,
  eds,  Astronomical Society of the Pacific Conference Series Vol. 281,
  Astronomical Data Analysis Software and Systems XI. p.~228

\bibitem[\protect\citeauthoryear{{Bloemen}, {Groot}, {Nelemans}  \&
  {Klein-Wolt}}{{Bloemen} et~al.}{2015}]{2015ASPC..496..254B}
{Bloemen} S.,  {Groot} P.,  {Nelemans} G.,   {Klein-Wolt} M.,  2015, in
  {Rucinski} S.~M.,  {Torres} G.,   {Zejda} M.,  eds,  Astronomical Society of
  the Pacific Conference Series Vol. 496, Living Together: Planets, Host Stars
  and Binaries. p.~254

\bibitem[\protect\citeauthoryear{{Boch}, {Fernique}, {Bonnarel}, {Chaitra},
  {Bot}, {Pineau}, {Baumann}  \& {Michel}}{{Boch} et~al.}{2020}]{hips2fits}
{Boch} T.,  {Fernique} P.,  {Bonnarel} F.,  {Chaitra} C.,  {Bot} C.,  {Pineau}
  F.~X.,  {Baumann} M.,   {Michel} L.,  2020, in {Pizzo} R.,  {Deul} E.~R.,
  {Mol} J.~D.,  {de Plaa} J.,   {Verkouter} H.,  eds,  Astronomical Society of
  the Pacific Conference Series Vol. 527, Astronomical Society of the Pacific
  Conference Series. p.~121

\bibitem[\protect\citeauthoryear{{Bradley} et~al.,}{{Bradley}
  et~al.}{2021}]{photutils}
{Bradley} L.,  et~al., 2021, {astropy/photutils: 1.1.0},
  \mn@doi{10.5281/zenodo.4624996}

\bibitem[\protect\citeauthoryear{{Brennan} \& {Fraser}}{{Brennan} \&
  {Fraser}}{2022}]{2022arXiv220102635B}
{Brennan} S.~J.,  {Fraser} M.,  2022, arXiv e-prints, \href
  {https://ui.adsabs.harvard.edu/abs/2022arXiv220102635B} {p. arXiv:2201.02635}

\bibitem[\protect\citeauthoryear{{Bulla}}{{Bulla}}{2019}]{2019MNRAS.489.5037B}
{Bulla} M.,  2019, \mn@doi [\mnras] {10.1093/mnras/stz2495}, \href
  {https://ui.adsabs.harvard.edu/abs/2019MNRAS.489.5037B} {489, 5037}

\bibitem[\protect\citeauthoryear{{Chornock} et~al.,}{{Chornock}
  et~al.}{2017}]{2017ApJ...848L..19C}
{Chornock} R.,  et~al., 2017, \mn@doi [\apjl] {10.3847/2041-8213/aa905c}, \href
  {https://ui.adsabs.harvard.edu/abs/2017ApJ...848L..19C} {848, L19}

\bibitem[\protect\citeauthoryear{{Coughlin} et~al.,}{{Coughlin}
  et~al.}{2018}]{CoDi2018}
{Coughlin} M.~W.,  et~al., 2018, \mn@doi [Monthly Notices of the Royal
  Astronomical Society] {10.1093/mnras/sty2174}, 480, 3871

\bibitem[\protect\citeauthoryear{Coughlin, Dietrich, Margalit  \&
  Metzger}{Coughlin et~al.}{2019}]{CoDi2018b}
Coughlin M.~W.,  Dietrich T.,  Margalit B.,   Metzger B.~D.,  2019, \mn@doi
  [Monthly Notices of the Royal Astronomical Society: Letters]
  {10.1093/mnrasl/slz133}, 489, L91

\bibitem[\protect\citeauthoryear{Coughlin et~al.,}{Coughlin
  et~al.}{2020a}]{PhysRevResearch.2.022006}
Coughlin M.~W.,  et~al., 2020a, \mn@doi [Phys. Rev. Research]
  {10.1103/PhysRevResearch.2.022006}, 2, 022006

\bibitem[\protect\citeauthoryear{{Coughlin} et~al.,}{{Coughlin}
  et~al.}{2020b}]{2020NatCo..11.4129C}
{Coughlin} M.~W.,  et~al., 2020b, \mn@doi [Nature Communications]
  {10.1038/s41467-020-17998-5}, \href
  {https://ui.adsabs.harvard.edu/abs/2020NatCo..11.4129C} {11, 4129}

\bibitem[\protect\citeauthoryear{{Coughlin} et~al.,}{{Coughlin}
  et~al.}{2020c}]{Coughlin2019}
{Coughlin} M.~W.,  et~al., 2020c, \mn@doi [\mnras] {10.1093/mnras/stz3457},
  \href {https://ui.adsabs.harvard.edu/abs/2020MNRAS.492..863C} {492, 863}

\bibitem[\protect\citeauthoryear{Coughlin et~al.,}{Coughlin
  et~al.}{2020d}]{implic2020}
Coughlin M.~W.,  et~al., 2020d, \mn@doi [Monthly Notices of the Royal
  Astronomical Society] {10.1093/mnras/staa1925}, 497, 1181–1196

\bibitem[\protect\citeauthoryear{{Coulter} et~al.,}{{Coulter}
  et~al.}{2017}]{2017Sci...358.1556C}
{Coulter} D.~A.,  et~al., 2017, \mn@doi [Science] {10.1126/science.aap9811},
  \href {http://adsabs.harvard.edu/abs/2017Sci...358.1556C} {358, 1556}

\bibitem[\protect\citeauthoryear{{Cowperthwaite} et~al.,}{{Cowperthwaite}
  et~al.}{2017}]{2017ApJ...848L..17C}
{Cowperthwaite} P.~S.,  et~al., 2017, \mn@doi [\apjl]
  {10.3847/2041-8213/aa8fc7}, \href
  {https://ui.adsabs.harvard.edu/abs/2017ApJ...848L..17C} {848, L17}

\bibitem[\protect\citeauthoryear{Cowperthwaite, Villar, Scolnic  \&
  Berger}{Cowperthwaite et~al.}{2019}]{CoVi2019}
Cowperthwaite P.~S.,  Villar V.~A.,  Scolnic D.~M.,   Berger E.,  2019, \mn@doi
  [The Astrophysical Journal] {10.3847/1538-4357/ab07b6}, 874, 88

\bibitem[\protect\citeauthoryear{{Dekany} et~al.,}{{Dekany}
  et~al.}{2020}]{2020PASP..132c8001D}
{Dekany} R.,  et~al., 2020, \mn@doi [\pasp] {10.1088/1538-3873/ab4ca2}, \href
  {https://ui.adsabs.harvard.edu/abs/2020PASP..132c8001D} {132, 038001}

\bibitem[\protect\citeauthoryear{{Dietrich}, {Coughlin}, {Pang}, {Bulla},
  {Heinzel}, {Issa}, {Tews}  \& {Antier}}{{Dietrich}
  et~al.}{2020}]{2020Sci...370.1450D}
{Dietrich} T.,  {Coughlin} M.~W.,  {Pang} P. T.~H.,  {Bulla} M.,  {Heinzel} J.,
   {Issa} L.,  {Tews} I.,   {Antier} S.,  2020, \mn@doi [Science]
  {10.1126/science.abb4317}, \href
  {https://ui.adsabs.harvard.edu/abs/2020Sci...370.1450D} {370, 1450}

\bibitem[\protect\citeauthoryear{Ducoin, Corre, Leroy  \& LeÂ Floch}{Ducoin
  et~al.}{2020}]{10.1093/mnras/staa114}
Ducoin J.-G.,  Corre D.,  Leroy N.,   LeÂ Floch E.,  2020, \mn@doi [Monthly
  Notices of the Royal Astronomical Society] {10.1093/mnras/staa114}, 492, 4768

\bibitem[\protect\citeauthoryear{Duverne et~al.,}{Duverne
  et~al.}{2022}]{muphoten}
Duverne P.~A.,  et~al., 2022, MUPHOTEN : a MUlti-band PHOtometry Tool for
  TElescope Network (\mn@eprint {arXiv} {2201.07565})

\bibitem[\protect\citeauthoryear{{Flaugher} et~al.,}{{Flaugher}
  et~al.}{2015}]{2015AJ....150..150F}
{Flaugher} B.,  et~al., 2015, \mn@doi [\aj] {10.1088/0004-6256/150/5/150},
  \href {https://ui.adsabs.harvard.edu/abs/2015AJ....150..150F} {150, 150}

\bibitem[\protect\citeauthoryear{{Ginsburg} et~al.,}{{Ginsburg}
  et~al.}{2019}]{astroquery}
{Ginsburg} A.,  et~al., 2019, \mn@doi [\aj] {10.3847/1538-3881/aafc33}, \href
  {http://adsabs.harvard.edu/abs/2019AJ....157...98G} {157, 98}

\bibitem[\protect\citeauthoryear{Goldstein et~al.,}{Goldstein
  et~al.}{2017}]{goldstein_ordinary_2017}
Goldstein A.,  et~al., 2017, \mn@doi [The Astrophysical Journal]
  {10.3847/2041-8213/aa8f41}, 848, L14

\bibitem[\protect\citeauthoryear{Gompertz et~al.,}{Gompertz
  et~al.}{2020}]{GOTO}
Gompertz B.~P.,  et~al., 2020, \mn@doi [Monthly Notices of the Royal
  Astronomical Society] {10.1093/mnras/staa1845}, 497, 726–738

\bibitem[\protect\citeauthoryear{{Graham} et~al.,}{{Graham}
  et~al.}{2019}]{2019PASP..131g8001G}
{Graham} M.~J.,  et~al., 2019, \mn@doi [\pasp] {10.1088/1538-3873/ab006c},
  \href {https://ui.adsabs.harvard.edu/abs/2019PASP..131g8001G} {131, 078001}

\bibitem[\protect\citeauthoryear{{Haggard}, {Nynka}, {Ruan}, {Kalogera},
  {Cenko}, {Evans}  \& {Kennea}}{{Haggard} et~al.}{2017}]{2017ApJ...848L..25H}
{Haggard} D.,  {Nynka} M.,  {Ruan} J.~J.,  {Kalogera} V.,  {Cenko} S.~B.,
  {Evans} P.,   {Kennea} J.~A.,  2017, \mn@doi [\apjl]
  {10.3847/2041-8213/aa8ede}, \href
  {https://ui.adsabs.harvard.edu/abs/2017ApJ...848L..25H} {848, L25}

\bibitem[\protect\citeauthoryear{{Hallinan} et~al.,}{{Hallinan}
  et~al.}{2017}]{Hallinan_2017Sci}
{Hallinan} G.,  et~al., 2017, \mn@doi [Science] {10.1126/science.aap9855},
  \href {https://ui.adsabs.harvard.edu/\#abs/2017Sci...358.1579H} {358, 1579}

\bibitem[\protect\citeauthoryear{{Ho} et~al.,}{{Ho}
  et~al.}{2022}]{Ho2022arXiv220112366H}
{Ho} A. Y.~Q.,  et~al., 2022, arXiv e-prints, \href
  {https://ui.adsabs.harvard.edu/abs/2022arXiv220112366H} {p. arXiv:2201.12366}

\bibitem[\protect\citeauthoryear{{Hotokezaka}, {Nakar}, {Gottlieb}, {Nissanke},
  {Masuda}, {Hallinan}, {Mooley}  \& {Deller}}{{Hotokezaka}
  et~al.}{2019}]{2019NatAs...3..940H}
{Hotokezaka} K.,  {Nakar} E.,  {Gottlieb} O.,  {Nissanke} S.,  {Masuda} K.,
  {Hallinan} G.,  {Mooley} K.~P.,   {Deller} A.~T.,  2019, \mn@doi [Nature
  Astronomy] {10.1038/s41550-019-0820-1}, \href
  {https://ui.adsabs.harvard.edu/abs/2019NatAs...3..940H} {3, 940}

\bibitem[\protect\citeauthoryear{{Hu} et~al.,}{{Hu}
  et~al.}{2017}]{2017SciBu..62.1433H}
{Hu} L.,  et~al., 2017, \mn@doi [Science Bulletin]
  {10.1016/j.scib.2017.10.006}, \href
  {https://ui.adsabs.harvard.edu/abs/2017SciBu..62.1433H} {62, 1433}

\bibitem[\protect\citeauthoryear{{Ivezi{\'c}} et~al.,}{{Ivezi{\'c}}
  et~al.}{2019}]{2019ApJ...873..111I}
{Ivezi{\'c}} {\v{Z}}.,  et~al., 2019, \mn@doi [\apj]
  {10.3847/1538-4357/ab042c}, \href
  {https://ui.adsabs.harvard.edu/abs/2019ApJ...873..111I} {873, 111}

\bibitem[\protect\citeauthoryear{{Karpov}}{{Karpov}}{2021}]{stdpipe}
{Karpov} S.,  2021, {STDPipe: Simple Transient Detection Pipeline} (\mn@eprint
  {ascl} {2112.006})

\bibitem[\protect\citeauthoryear{Kasen, Metzger, Barnes, Quataert  \&
  Ramirez-Ruiz}{Kasen et~al.}{2017}]{2017}
Kasen D.,  Metzger B.,  Barnes J.,  Quataert E.,   Ramirez-Ruiz E.,  2017,
  \mn@doi [Nature] {10.1038/nature24453}, 551, 80–84

\bibitem[\protect\citeauthoryear{Kasliwal et~al.,}{Kasliwal
  et~al.}{2020}]{Kasliwal_2020}
Kasliwal M.~M.,  et~al., 2020, \mn@doi [The Astrophysical Journal]
  {10.3847/1538-4357/abc335}, 905, 145

\bibitem[\protect\citeauthoryear{{Kostov} \& {Bonev}}{{Kostov} \&
  {Bonev}}{2018}]{ps1_to_stetson}
{Kostov} A.,  {Bonev} T.,  2018, Bulgarian Astronomical Journal, \href
  {https://ui.adsabs.harvard.edu/abs/2018BlgAJ..28....3K} {28, 3}

\bibitem[\protect\citeauthoryear{{Lang}, {Hogg}, {Mierle}, {Blanton}  \&
  {Roweis}}{{Lang} et~al.}{2010}]{2010astrometrynet}
{Lang} D.,  {Hogg} D.~W.,  {Mierle} K.,  {Blanton} M.,   {Roweis} S.,  2010,
  \mn@doi [\aj] {10.1088/0004-6256/139/5/1782}, 139, 1782

\bibitem[\protect\citeauthoryear{{Levan} et~al.,}{{Levan}
  et~al.}{2005}]{LeNu2005}
{Levan} A.,  et~al., 2005, \mn@doi [\apj] {10.1086/428657}, \href
  {https://ui.adsabs.harvard.edu/abs/2005ApJ...624..880L} {624, 880}

\bibitem[\protect\citeauthoryear{{Margalit} \& {Metzger}}{{Margalit} \&
  {Metzger}}{2017}]{2017ApJ...850L..19M}
{Margalit} B.,  {Metzger} B.~D.,  2017, \mn@doi [\apjl]
  {10.3847/2041-8213/aa991c}, \href
  {https://ui.adsabs.harvard.edu/abs/2017ApJ...850L..19M} {850, L19}

\bibitem[\protect\citeauthoryear{Margutti et~al.,}{Margutti
  et~al.}{2018}]{margutti2018target}
Margutti R.,  et~al., 2018, Target of Opportunity Observations of Gravitational
  Wave Events with LSST (\mn@eprint {arXiv} {1812.04051})

\bibitem[\protect\citeauthoryear{{Masci} et~al.,}{{Masci}
  et~al.}{2019}]{2019PASP..131a8003M}
{Masci} F.~J.,  et~al., 2019, \mn@doi [\pasp] {10.1088/1538-3873/aae8ac}, \href
  {https://ui.adsabs.harvard.edu/abs/2019PASP..131a8003M} {131, 018003}

\bibitem[\protect\citeauthoryear{{McCully} \& {Tewes}}{{McCully} \&
  {Tewes}}{2019}]{astroscrappy}
{McCully} C.,  {Tewes} M.,  2019, {Astro-SCRAPPY: Speedy Cosmic Ray
  Annihilation Package in Python} (\mn@eprint {ascl} {1907.032})

\bibitem[\protect\citeauthoryear{{Morgan}, {Kaiser}, {Moreau}, {Anderson}  \&
  {Burgett}}{{Morgan} et~al.}{2012}]{2012SPIE.8444E..0HM}
{Morgan} J.~S.,  {Kaiser} N.,  {Moreau} V.,  {Anderson} D.,   {Burgett} W.,
  2012, in {Stepp} L.~M.,  {Gilmozzi} R.,   {Hall} H.~J.,  eds,  Society of
  Photo-Optical Instrumentation Engineers (SPIE) Conference Series Vol. 8444,
  Ground-based and Airborne Telescopes IV. p. 84440H (\mn@eprint {arXiv}
  {1207.2513}), \mn@doi{10.1117/12.926646}

\bibitem[\protect\citeauthoryear{Möller et~al.,}{Möller
  et~al.}{2020}]{10.1093/mnras/staa3602}
Möller A.,  et~al., 2020, \mn@doi [Monthly Notices of the Royal Astronomical
  Society] {10.1093/mnras/staa3602}, 501, 3272

\bibitem[\protect\citeauthoryear{{Patterson} et~al.,}{{Patterson}
  et~al.}{2019}]{2019PASP..131a8001P}
{Patterson} M.~T.,  et~al., 2019, \mn@doi [\pasp] {10.1088/1538-3873/aae904},
  \href {https://ui.adsabs.harvard.edu/abs/2019PASP..131a8001P} {131, 018001}

\bibitem[\protect\citeauthoryear{{Perley} et~al.,}{{Perley}
  et~al.}{2019}]{2019MNRAS.484.1031P}
{Perley} D.~A.,  et~al., 2019, \mn@doi [\mnras] {10.1093/mnras/sty3420}, \href
  {https://ui.adsabs.harvard.edu/abs/2019MNRAS.484.1031P} {484, 1031}

\bibitem[\protect\citeauthoryear{{Petrov} et~al.,}{{Petrov}
  et~al.}{2022}]{2022ApJ...924...54P}
{Petrov} P.,  et~al., 2022, \mn@doi [\apj] {10.3847/1538-4357/ac366d}, \href
  {https://ui.adsabs.harvard.edu/abs/2022ApJ...924...54P} {924, 54}

\bibitem[\protect\citeauthoryear{Piro, Haynie  \& Yao}{Piro
  et~al.}{2021}]{PiHa2021}
Piro A.~L.,  Haynie A.,   Yao Y.,  2021, \mn@doi [The Astrophysical Journal]
  {10.3847/1538-4357/abe2b1}, 909, 209

\bibitem[\protect\citeauthoryear{Savchenko et~al.,}{Savchenko
  et~al.}{2017}]{savchenko_integral_2017}
Savchenko V.,  et~al., 2017, \mn@doi [The Astrophysical Journal]
  {10.3847/2041-8213/aa8f94}, 848, L15

\bibitem[\protect\citeauthoryear{{Tanvir}, {Levan}, {Fruchter}, {Hjorth},
  {Hounsell}, {Wiersema}  \& {Tunnicliffe}}{{Tanvir}
  et~al.}{2013}]{2013Natur.500..547T}
{Tanvir} N.~R.,  {Levan} A.~J.,  {Fruchter} A.~S.,  {Hjorth} J.,  {Hounsell}
  R.~A.,  {Wiersema} K.,   {Tunnicliffe} R.~L.,  2013, \mn@doi [\nat]
  {10.1038/nature12505}, \href
  {https://ui.adsabs.harvard.edu/abs/2013Natur.500..547T} {500, 547}

\bibitem[\protect\citeauthoryear{{Tonry} et~al.,}{{Tonry}
  et~al.}{2018}]{2018PASP..130f4505T}
{Tonry} J.~L.,  et~al., 2018, \mn@doi [\pasp] {10.1088/1538-3873/aabadf}, \href
  {https://ui.adsabs.harvard.edu/abs/2018PASP..130f4505T} {130, 064505}

\bibitem[\protect\citeauthoryear{{Troja} et~al.,}{{Troja}
  et~al.}{2018}]{TrPi2018}
{Troja} E.,  et~al., 2018, \mn@doi [Monthly Notices of the Royal Astronomical
  Society] {10.1093/mnrasl/sly061}, \href
  {http://adsabs.harvard.edu/abs/2018MNRAS.tmpL..60T} {}

\bibitem[\protect\citeauthoryear{{Waters} et~al.,}{{Waters}
  et~al.}{2020}]{ps1images}
{Waters} C.~Z.,  et~al., 2020, \mn@doi [\apjs] {10.3847/1538-4365/abb82b},
  \href {https://ui.adsabs.harvard.edu/abs/2020ApJS..251....4W} {251, 4}

\bibitem[\protect\citeauthoryear{{van Dokkum}}{{van Dokkum}}{2001}]{lacosmic}
{van Dokkum} P.~G.,  2001, \mn@doi [\pasp] {10.1086/323894}, \href
  {https://ui.adsabs.harvard.edu/abs/2001PASP..113.1420V} {113, 1420}

\makeatother
\end{thebibliography}

\appendix

\section{Observations from the GRANDMA program}
\label{obsApp}


\begin{table*}
\caption{Upper limits - Summary of the GRANDMA observations of some KN-MANGROVE alerts. $\delta \mathrm{t}$ is the delay between the beginning of the observation and the public detection discovery. In this table, only \textbf{upper limits} useful to confirm the nature of the source as moving objects are reported. Magnitudes are given in the AB system  and not correct for Galactic extinction, upper limits are given at $5\sigma$ confidence.}
\label{tab:KN_upper}
\begin{tabular}{cccccccccl}
Source &  Obs date & Time & $\delta \mathrm{t}$ (days) & filter & Detec./Upper (mag) & Telescope/Observer & Reduction pipeline \\
\hline
ZTF21abdwdwo & 2021-06-04 & 04:27:26 &    0.0   & $r^\prime$   & $18.8\pm0.1$ &                ZTF & ZTF\\
ZTF21abdwdwo & 2021-06-05 & 00:00:00 &    0.8   & Clear   & $>20.7$ & T40-A77DAU & Muphoten \\
ZTF21abdwdwo & 2021-06-07 & 21:41:04 &    3.7   & $B$   & $>21.4$ & T-CAT & Muphoten \\
\hline
ZTF21abxkven & 2021-09-03 & 08:28:07 & 0.0 & $r^\prime$ & $18.4\pm0.1$ & ZTF & ZTF \\
ZTF21abxkven & 2021-09-03 & 22:13:14 & 0.5 & $B$ & $>19.0$ & Abastumani-T70 & Muphoten \\
ZTF21abxkven & 2021-09-03 & 22:14:28 & 0.5 & $R_C$ & $>18.8$ & Abastumani-T70 & Muphoten \\
ZTF21abxkven & 2021-09-03 & 22:35:46 & 0.5 & $L$ & $>18.2$ & K26 & Muphoten \\
ZTF21abxkven & 2021-09-04 & 02:42:20 & 0.7 & $L$ & $>20.6$ & T-STSOPHIE & Muphoten \\
ZTF21abxkven & 2021-09-04 & 13:07:55 & 1.6 & Clear & $>20.6$ & T40-A77DAU & Muphoten \\
\hline
ZTF21abxlpdl & 2021-09-03 &  08:59:16 & 0.0 & $r^\prime$  & $19.3\pm0.1$ & ZTF & ZTF \\
ZTF21abxlpdl & 2021-09-03 & 20:56:08 & 0.5 & $B$  & $>19.1$ & Abastumani-T70 & Muphoten \\
ZTF21abxlpdl & 2021-09-03 & 20:57:22 & 0.5 & $R_C$  & $>19.1$ & Abastumani-T70 & Muphoten \\
ZTF21abxlpdl & 2021-09-04 & 23:32:00 & 1.6 & Clear  & $>20.7$ & T40-A77DAU & Muphoten \\
\hline
ZTF21abyplur & 2021-09-06 & 09:12:27 & 0.0 & $r^\prime$  & $17.5\pm0.1$ & ZTF & ZTF \\
ZTF21abyplur & 2021-09-15 & 21:17:03 & 9.5 & $g^\prime$  & $>$ 18.3 & Tibet-50 & Muphoten \\
\end{tabular}
\end{table*}

\begin{table*}
\caption{Detections - Summary of the GRANDMA observations of \ztfink{} candidates. $\delta \mathrm{t}$ is the delay between the beginning of the observation and the public detection discovery. In this table, only detection magnitudes of each observational epoch are reported. Magnitudes are given in the AB system and not correct for Galactic extinction.}
\label{tab:KN_observations}
\begin{tabular}{cccccccccl}
Source &  Obs date & Time & $\delta \mathrm{t}$ (days) & filter & Detec. mag (STDpipe) & mag. Muphoten (or ZTF) & Telescope/Observer & \\
\hline

ZTF21abbzjeq  & 2021-05-20  & 09:28:49 & 0.0   & $g^\prime$   & -                  & $19.7\pm0.2$   & ZTF  & \\
ZTF21abbzjeq  &2021-05-27  & 07:35:33 & 6.9   & $g^\prime$   & -                  & $17.5\pm0.1$   & ZTF  & \\
ZTF21abbzjeq  & 2021-05-27  & 09:12:08 & 7.0   & $r^\prime$   & -                  & $17.7\pm0.1$   & ZTF  & \\
ZTF21abbzjeq & 2021-05-28 & 22:14:18 & 8.5  & $V$  & $17.5\pm0.1$ &  $17.5\pm0.1$ & T-BRO & \\
ZTF21abbzjeq & 2021-05-28 & 22:29:27 & 8.5  & $R$  & $17.4\pm0.1$ &  $17.4\pm0.1$ & T-BRO & \\
ZTF21abbzjeq & 2021-05-28 & 22:49:12 & 8.6  & $B$  & $17.4\pm0.1$ &  $17.4\pm0.1$ & T-BRO & \\
ZTF21abbzjeq & 2021-05-28 & 23:10:23 & 8.6  & $I$  & $17.5\pm0.1$ &  $17.4\pm0.1$ & T-BRO & \\
ZTF21abbzjeq  & 2021-05-29  & 06:44:06 & 8.9   & $g^\prime$   & -                  & $17.3\pm0.1$   & ZTF  & \\
ZTF21abbzjeq  & 2021-05-31  & 08:25:51 & 11.0   & $r^\prime$   & -                  & $17.3\pm0.1$   & ZTF  & \\
ZTF21abbzjeq & 2021-05-31 & 22:53:44 & 11.6  & $R$  & $17.2\pm0.1$ &  $17.2\pm0.1$ & Omegon203 & \\
ZTF21abbzjeq  & 2021-06-02  & 07:10:42 & 12.9   & $g^\prime$   & -                  & $17.1\pm0.1$   & ZTF  & \\
ZTF21abbzjeq  & 2021-08-01   & 06:17:15 & 72.9   & $g^\prime$   & -                  & $20.1\pm0.1$   & ZTF  & \\
ZTF21abbzjeq & 2021-08-03 & 00:44:35 &  74.6     & $R$   & $19.1\pm0.1$ &     $19.0\pm0.1$        & VIRT &\\ 
ZTF21abbzjeq & 2021-08-03 & 17:58:53 &  75.3      & $B$   &        $20.1\pm0.1$            & $20.3\pm0.1$ &  Abastumani/T70 &\\ 
ZTF21abbzjeq & 2021-08-03 & 18:01:07 &   75.4    & $R_C$ &         $18.9\pm0.1$            &  $18.9\pm0.1$   &  Abastumani/T70 &\\ 
ZTF21abbzjeq & 2021-08-04 & 04:40:50 & 75.8 & $R$   & $19.3\pm0.2$   &  $18.9\pm0.2$ & TRT-SRO &\\ 
ZTF21abbzjeq & 2021-08-04 & 06:23:46 & 75.9 & $I$   & $19.2\pm0.2$   &  $19.4\pm0.2$ & TRT-SRO &\\ 
ZTF21abbzjeq  & 2021-08-08  & 04:29:20 & 79.8   & $r^\prime$   & -                  & $19.6\pm0.2$   & ZTF  & \\
ZTF21abbzjeq  & 2021-08-08  & 05:40:55 & 79.8   & $g^\prime$   & -                  & $20.2\pm0.3$   & ZTF  & \\
\hline
ZTF21abfaohe & 2021-06-09  & 05:14:26  & 0.0  & $g^\prime$  & - & $19.4\pm0.2$  & ZTF  & \\
ZTF21abfaohe & 2021-06-22  & 04:32:33  & 13.0  & $r^\prime$  & - & $14.4\pm0.2$  & ZTF  & \\
ZTF21abfaohe & 2021-06-22  & 07:02:41  & 13.1  & $g^\prime$  & - & $14.3\pm0.2$  & ZTF  & \\
ZTF21abfaohe & 2021-06-25 & 20:23:39 &   16.6    & $r^\prime$   &  $14.2\pm0.1$ &  $14.2\pm0.2$   & Iris &\\
ZTF21abfaohe & 2021-06-25 & 21:02:47 &   16.7    & $g^\prime$   &  $14.1\pm0.1$ & $14.1\pm0.2$    & Iris &\\
ZTF21abfaohe & 2021-06-25 & 21:23:27 &   16.7    & $G$   &  $14.2\pm0.1$ &  -   & PDAObs &\\
ZTF21abfaohe & 2021-06-25 & 22:01:28 &   16.7    & $B$  &  $14.2\pm0.1$ &  $14.5\pm0.1$    &  T-CAT &\\
ZTF21abfaohe & 2021-06-25 & 22:01:28 &   16.7    & $G$  &  $14.2\pm0.1$ &   $14.4\pm0.1$  &  T-CAT &\\
ZTF21abfaohe & 2021-06-25 & 22:01:28 &   16.7    & $R$  &  $14.1\pm0.1$ &   $13.9\pm0.1$  &  T-CAT &\\
ZTF21abfaohe & 2021-06-25 & 22:10:50 &   16.7    & $R$   &  $14.2\pm0.1$ &    -   & PDAObs &\\
ZTF21abfaohe & 2021-06-25 & 22:58:41 &   16.8    & $G$   &  $14.2\pm0.1$ &  -   & PDAObs &\\
ZTF21abfaohe & 2021-06-26 & 00:00:00 &   16.8    & Clear  &  $14.1\pm0.1$ &  $13.5\pm0.1$   & T40-A77DAU &\\
ZTF21abfaohe & 2021-06-26  & 05:07:59  & 17.0  & $g^\prime$  & - & $14.1\pm0.1$  & ZTF  & \\
ZTF21abfaohe & 2021-06-26  & 06:32:45  & 17.1  & $r^\prime$  & - & $14.3\pm0.1$  & ZTF  & \\
ZTF21abfaohe & 2021-06-26 & 20:18:11 &   17.6    & Clear   &  $14.1\pm0.1$ & -    & MSXD-A77 &\\
ZTF21abfaohe & 2021-06-26 & 20:26:08 &   17.7    & $R$   &  $14.2\pm0.1$ & -   & MSXD-A77 &\\
ZTF21abfaohe & 2021-06-27 & 21:42:45 &   18.7    & Clear  &  $14.1\pm0.1$ &  $13.6\pm0.1$    & T40-A77DAU &\\
ZTF21abfaohe & 2021-06-28  & 05:04:28  & 19.0  & $r^\prime$  & - & $14.3\pm0.1$  & ZTF  & \\
ZTF21abfaohe & 2021-06-28  & 07:02:23  & 19.1  & $g^\prime$  & - & $14.2\pm0.1$  & ZTF  & \\
ZTF21abfaohe & 2021-06-28 & 09:11:07 &   19.2    & $r^\prime$  &  $14.2\pm0.1$ &  $14.3\pm0.1$   & iTel-17 &\\
ZTF21abfaohe & 2021-06-28 & 09:21:10 &   19.2    & $g^\prime$  &  $14.1\pm0.1$ & $14.2\pm0.1$    & iTel-17 &\\
ZTF21abfaohe & 2021-06-30  & 05:35:16  & 21.0  & $g^\prime$  & - & $14.2\pm0.1$  & ZTF  & \\
ZTF21abfaohe & 2021-07-01 & 21:32:09 &   22.7    & Clear  &  $14.2\pm0.1$ &   $13.7\pm0.1$  & T40-A77DAU &\\
ZTF21abfaohe & 2021-07-02  & 05:34:58  & 23.0  & $g^\prime$  & - & $14.2\pm0.1$  & ZTF  & \\
ZTF21abfaohe & 2021-07-04 & 07:19:38 &   25.1    & Clear  &  $14.3\pm0.1$ &  $13.9\pm0.3$   & Beverly-Begg &\\
ZTF21abfaohe & 2021-07-04 & 23:02:45 &   25.8    & $B$  &  $14.5\pm0.1$ & $14.9\pm0.1$     &  T-CAT &\\
ZTF21abfaohe & 2021-07-04 & 23:02:45 &   25.8    & $G$  &  $14.4\pm0.1$ &  $14.7\pm0.1$   &  T-CAT &\\
ZTF21abfaohe & 2021-07-04 & 23:02:45 &   25.8    & $R$  &  $14.4\pm0.1$ &  $14.2\pm0.1$   &  T-CAT &\\
ZTF21abfaohe & 2021-07-05  & 04:34:55  & 26.0  & $r^\prime$  & - & $14.5\pm0.1$  & ZTF  & \\
ZTF21abfaohe & 2021-07-05  & 06:05:02  & 26.1  & $g^\prime$  & - & $14.4\pm0.1$  & ZTF  & \\
ZTF21abfaohe & 2021-07-07  & 05:32:50  & 27.0  & $r^\prime$  & - & $14.7\pm0.1$  & ZTF  & \\
ZTF21abfaohe & 2021-07-08 & 20:34:29 &   29.7    & Clear  &  $14.7\pm0.1$ & $14.3\pm0.1$    &  ZnithObs &\\
ZTF21abfaohe & 2021-07-09  & 05:29:28  & 30.0  & $r^\prime$  & - & $14.9\pm0.1$  & ZTF  & \\
ZTF21abfaohe & 2021-07-28  & 04:38:47  & 49.0  & $g^\prime$  & - & $16.5\pm0.1$  & ZTF  & \\
ZTF21abfaohe & 2021-07-31 & 00:46:46 &  52.8     & $R_C$ &     $15.5\pm0.1$               &     -   & FRAM-Auger &\\
ZTF21abfaohe & 2021-08-01  & 04:16:28  & 53.0  & $r^\prime$  & - & $15.8\pm0.1$  & ZTF  & \\
ZTF21abfaohe & 2021-08-03 & 17:21:26 &   55.5    & $B$   &  $17.1\pm0.1$  & $17.1\pm0.1$  &  Abastumani/T70 & \\
ZTF21abfaohe & 2021-08-03 & 17:22:40 &   55.5    & $R_C$ &  $15.9\pm0.1$ &  $15.9\pm0.1$ &  Abastumani/T70 \\
\hline
ZTF21abfmbix & 2021-06-11 & 05:14:49 & 0.0   & $g^\prime$   & -                  & $18.1\pm0.10$   & ZTF  \\
ZTF21abfmbix & 2021-06-11 & 06:05:58 & 0.1 & $r^\prime$  & - & $17.8\pm0.10$ & ZTF &   & \\
ZTF21abfmbix & 2021-06-11 & 21:09:01  & 0.7 & $L$ & $17.4\pm0.1$ &  $16.4\pm0.2$ & Uranoscope  \\
ZTF21abfmbix & 2021-06-11 & 21:21:33  & 0.7 & $R$ & $17.4\pm0.1$ &  $17.2\pm0.2$ & MSXD-A77 &   \\
ZTF21abfmbix & 2021-06-11 &  21:41:37 & 0.7  & L & $17.0\pm0.1$  & $16.8\pm0.2$ & Teams&    \\
ZTF21abfmbix & 2021-06-11 &  21:54:07 & 0.7 & $R$ & $17.4\pm0.1$ & $17.2\pm0.1$ & T-CAT &   \\
\end{tabular}
\end{table*}

\begin{table*}
\addtocounter{table}{-1}
\caption{Continued.}
\begin{tabular}{cccccccccl}
Source &  Obs date & Time & $\delta \mathrm{t}$ (days) & filter & Detec. mag (STDpipe) & Muphoten (or ZTF) & Telescope/Observer & \\
\hline
ZTF21abfmbix & 2021-06-11 &  21:54:07 & 0.7  & $B$ & $17.8\pm0.1$  & $18.1\pm0.1$ & T-CAT &     \\
ZTF21abfmbix & 2021-06-11 &  21:54:07 & 0.7 & $G$ & $17.5\pm0.1$  & $17.5\pm0.1$ & T-CAT &    \\
ZTF21abfmbix & 2021-06-11 &  22:20:06 & 0.7  & $R$ & $17.4\pm0.1$   & $17.4\pm0.2$  & Vallieres   & \\
ZTF21abfmbix & 2021-06-12 &  21:20:54 & 1.7   & Clear & $17.1\pm0.2$  & $16.9\pm0.2$      & T-GRA   & \\
ZTF21abfmbix & 2021-06-12 &  21:42:53  & 1.7 & $R$   & $17.0\pm0.1$  & $16.6\pm0.1$  & T-CAT     & \\
ZTF21abfmbix & 2021-06-12 &  21:42:53 &  1.7  & $B$  & $17.2\pm0.1$  & $17.6\pm0.1$  & T-CAT    & \\
ZTF21abfmbix & 2021-06-12 &21:42:53 & 1.7 & $G$  & $17.1\pm0.1$ & $17.4\pm0.1 $ & T-CAT    & \\
ZTF21abfmbix & 2021-06-12 &21:46:29 &  1.7  & $L$ & $17.0\pm0.1$   & $16.8\pm0.2$      & N250-ROU     & \\
ZTF21abfmbix & 2021-06-13 &02:56:45  & 1.9 & Clear  & $16.9\pm0.1$     & $16.5\pm0.5 $  & RIT   & \\
ZTF21abfmbix &  2021-06-13 &04:25:03  & 2.0 & $G$  &  $17.0\pm0.1$  & $17.0\pm0.1$  & C11FREE    & \\
ZTF21abfmbix &  2021-06-13 &04:25:03  & 2.0  & $R$    & $16.9\pm0.1$  & $16.9\pm0.2$  & C11FREE &   & \\
ZTF21abfmbix & 2021-06-13 &04:38:39  &  2.0 & $g^\prime$    & - & $17.1\pm0.1$  & ZTF &   & \\
ZTF21abfmbix & 2021-06-13 &05:43:19  & 2.0  & $V$ & $17.0\pm0.1$ & $17.0\pm0.1$   & iTel-24  &   & \\
ZTF21abfmbix & 2021-06-13 &05:51:15  & 2.0  & $I$ & $17.0\pm0.2$ & $17.1\pm0.1$   & iTel-24  &   & \\
ZTF21abfmbix & 2021-06-13& 05:59:57  &  2.0 & $r^\prime$    & - & $16.9\pm0.1$  & ZTF &   & \\
ZTF21abfmbix & 2021-06-13 &21:53:06  & 2.7 & $R$     & $16.6\pm0.1$ &  $16.4\pm0.1$  & T-CAT  &   & \\
ZTF21abfmbix &  2021-06-13& 21:53:06 & 2.7   & $B$    & $16.9\pm0.1$ & $17.2\pm0.1$   & T-CAT  &   & \\
ZTF21abfmbix & 2021-06-13 &21:53:06  & 2.7   & $G$    & $16.8\pm0.1$ & $16.9\pm0.1$   & T-CAT  &   & \\
ZTF21abfmbix & 2021-06-13 &22:00:44 &  2.7  &$R$ & $16.7\pm0.1$  & $16.5\pm0.1$    & Omegon23 &   & \\
ZTF21abfmbix & 2021-07-01 &04:29:46 & 20.0   & $r^\prime$ &       & $15.5\pm0.1$  & ZTF   & & \\
ZTF21abfmbix & 2021-07-01& 05:59:37 & 20.0    & $g^\prime$ &      & $16.2\pm0.1$ & ZTF   & \\
ZTF21abfmbix & 2021-07-04& 22:35:05 & 23.7    & $R$ & $15.7\pm0.1$  & $15.5\pm0.1$    & T-CAT  &   & \\
ZTF21abfmbix & 2021-07-04& 22:35:05 & 23.7   & $B$ &   $16.7\pm0.1$     & $17.1\pm0.1$ & T-CAT   & \\
ZTF21abfmbix & 2021-07-04& 22:35:05 &  23.7  & $G$ & $16.1\pm0.1$  & $16.2\pm0.1$     & T-CAT  &   & \\
ZTF21abfmbix &  2021-07-10& 21:21:06 & 29.7    & Clear & $16.2\pm0.1$  & $15.2\pm0.1$      & T40-A77DAU &   & \\
\hline
ZTF21ablssud & 2021-07-14 & 07:06:24 & 0.0   & $r^\prime$   & -                  & $16.4\pm0.1$   & ZTF  \\
ZTF21ablssud & 2021-07-14 & 08:00:27 & 0.0   & $g^\prime$   & -                  & $16.4\pm0.1$   & ZTF  \\
ZTF21ablssud & 2021-07-16 & 06:25:45 & 2.0   & $r^\prime$   & -                  & $16.7\pm0.1$   & ZTF  \\
ZTF21ablssud & 2021-07-16 & 07:34:52 & 2.0   & $g^\prime$   & -                  & $16.9\pm0.1$   & ZTF  \\
ZTF21ablssud & 2021-07-16 & 21:11:45  & 2.6   & $G$ & $17.0\pm0.1$    &  $16.9\pm0.1$ & PDAObs &\\
ZTF21ablssud & 2021-07-16 & 21:34:49  & 2.6   & $V$ & $17.0\pm0.1$    & $17.0\pm0.1$ & Gallinero &\\
ZTF21ablssud & 2021-07-16 & 21:39:48  & 2.6   & $G$ & $16.8\pm0.1$    & $17.4\pm0.1$ & T-CAT &\\
ZTF21ablssud & 2021-07-16 & 21:39:48  & 2.6   & $B$ & $17.1\pm0.1$    & $17.7\pm0.1$ & T-CAT &\\
ZTF21ablssud & 2021-07-16 & 21:39:48  & 2.6   & $R$ & $16.7\pm0.1$    & $16.6\pm0.1$ & T-CAT &\\
ZTF21ablssud & 2021-07-16 & 21:42:04  & 2.6   & $R$ & $16.8\pm0.1$    & $16.5\pm0.1$ & PDAObs &\\
ZTF21ablssud & 2021-07-16 & 22:12:09  & 2.6   & $I$ & $16.5\pm0.1$    & $16.6\pm0.1$ & PDAObs &\\
ZTF21ablssud & 2021-07-16 & 22:18:17  & 2.6   & $L$ & $16.7\pm0.1$    & $16.1\pm0.1$ & Uranoscope &\\
ZTF21ablssud & 2021-07-16 & 22:24:00  & 2.6   & Clear & $16.7\pm0.1$    & $16.0 \pm$ 0.1 & MSXD-A77 &\\
ZTF21ablssud & 2021-07-17 & 21:21:24  & 3.6   & Clear & $16.9\pm0.1$    & $16.4\pm0.1$ & T40-A77DAU &\\
ZTF21ablssud & 2021-07-17 & 21:34:01  & 3.6   & $V$ & $16.9\pm0.1$    & $16.8\pm0.1$ & Omegon203 &\\
ZTF21ablssud & 2021-07-17 & 21:35:18  & 3.6   & $V$ & $17.2\pm0.1$    & $17.2\pm0.1$ &  Gallinero &\\
ZTF21ablssud & 2021-07-17 & 21:48:52  & 3.6   & $R$ & $17.0\pm0.1$    & $16.8\pm0.2$ & N250-ROU &\\
ZTF21ablssud & 2021-07-17 & 21:59:42  & 3.6   & $R$ & $16.9\pm0.1$    & $16.7\pm0.2$ & Montarrenti &\\
ZTF21ablssud & 2021-07-17 & 22:01:39  & 3.6   & $B$ & $17.4\pm0.1$    & $17.6\pm0.2$ & N250-ROU &\\
ZTF21ablssud & 2021-07-17 & 22:15:59  & 3.6   & $G$ & $17.1\pm0.1$    & $17.3\pm$ 0.2 & N250-ROU &\\
ZTF21ablssud & 2021-07-17 & 23:16:15  & 3.7   & $R$ & $17.0\pm0.1$    & $16.9\pm0.2$ & N250-ROU &\\
ZTF21ablssud & 2021-07-17 & 23:22:48  & 3.7   & $B$ & $17.3\pm0.1$    & $17.6\pm0.2$ & N250-ROU &\\
ZTF21ablssud & 2021-07-17 & 23:36:58  & 3.7   & $B$ & $17.4\pm0.1$    & $17.5\pm0.2$ & Montarrenti &\\
ZTF21ablssud & 2021-07-17 & 23:36:58  & 3.7   & $R$ & $16.9\pm0.1$    & $17.0\pm0.2$ & Montarrenti &\\
ZTF21ablssud & 2021-07-18 & 01:50:55  & 3.8   & $B$ & $17.6\pm0.2$    & $17.6\pm0.1$ & T-BRO  &\\
ZTF21ablssud & 2021-07-18 & 02:03:09  & 3.8   & $I$ & $16.7\pm0.1$    & $16.6\pm0.1$ & T-BRO  &\\
ZTF21ablssud & 2021-07-18 & 04:37:08  & 3.9   & $R$ & $17.1\pm0.1$    & $16.8\pm0.2$ & C11FREE  &\\
ZTF21ablssud & 2021-07-18 & 04:37:08  & 3.9   & $G$ & $17.5\pm0.1$    & $17.4\pm0.1$ & C11FREE  &\\
ZTF21ablssud & 2021-07-18 & 21:09:46  & 4.6   & $g^\prime$ & $17.2\pm0.1$    & $17.1\pm0.1$ & PDAObs  &\\
ZTF21ablssud & 2021-07-18 & 21:23:05  & 4.6   & $R$ & $16.8\pm0.1$    & $17.0\pm0.3$ & Vallieres  &\\
ZTF21ablssud & 2021-07-18 & 21:29:49  & 4.6   & Clear & $16.9\pm0.1$    & $16.4\pm0.1$ & T40-A77DAU  &\\
ZTF21ablssud & 2021-07-18 & 21:40:09  & 4.6   & $G$ & $17.1\pm0.1$    & $17.7\pm0.1$ & T-CAT  &\\
ZTF21ablssud & 2021-07-18 & 21:40:09  & 4.6   & $R$ & $17.0\pm0.1$    &  $16.8\pm0.1$ & T-CAT  &\\
ZTF21ablssud & 2021-07-18 & 21:40:09  & 4.6   & $B$ & $17.4\pm0.1$    &  $18.0\pm0.1$ & T-CAT  &\\
ZTF21ablssud & 2021-07-18 & 21:55:21  & 4.6   & $r^\prime$ & $17.1\pm0.1$    & $17.1\pm0.1$ & PDAObs  &\\
ZTF21ablssud & 2021-07-18 & 22:41:05  & 4.6   & $i$ & $16.8\pm0.1$    & $16.9\pm0.1$ & PDAObs  &\\
ZTF21ablssud & 2021-07-19 & 01:45:49  & 4.8   & $B$ & $17.6\pm0.1$    & $17.6\pm0.1$ & T-BRO  &\\
ZTF21ablssud & 2021-07-19 & 01:54:56  & 4.8   & $V$ & $17.2\pm0.1$    & $17.3\pm0.1$ & T-BRO  &\\
ZTF21ablssud & 2021-07-19 & 02:04:03  & 4.8   & $R$ & $17.0\pm0.1$    & $16.8\pm0.1$ & T-BRO  &\\
ZTF21ablssud & 2021-07-19 & 02:13:08  & 4.8   & $I$ & $16.9\pm0.1$    & $16.6\pm0.1$ & T-BRO  &\\
\end{tabular}
\end{table*}

\begin{table*}
\addtocounter{table}{-1}
\caption{Continued.}
\begin{tabular}{cccccccccl}
Source &  Obs date & Time & $\delta \mathrm{t}$ (days) & filter & Detec. mag (STDpipe) & Muphoten (or ZTF) & Telescope/Observer & \\
\hline
ZTF21ablssud & 2021-07-19 & 07:06:07  & 5.0   & $r^\prime$   & -                  & $17.2\pm0.1$   & ZTF  \\
ZTF21ablssud & 2021-07-19 & 21:02:35  & 5.6   & $Clear$ & $17.2\pm0.1$    & $16.6\pm0.1$ & T40-A77DAU  &\\
ZTF21ablssud & 2021-07-19 & 21:23:50  & 5.6   & $R$ & $17.1\pm0.1$    & $17.0\pm0.2$ & Montarrenti  &\\
ZTF21ablssud & 2021-07-19 & 22:12:46  & 5.6   & $L$ & $17.3\pm0.1$    & $16.4\pm0.1$ & K26  &\\
ZTF21ablssud & 2021-07-19 & 22:19:51  & 5.6   & $R$ & $17.2\pm0.1$    & $17.0\pm0.1$ & T-CAT  &\\
ZTF21ablssud & 2021-07-19 & 22:19:51  & 5.6   & $B$ & $17.6\pm0.1$    & $18.2\pm0.1$ & T-CAT  &\\
ZTF21ablssud & 2021-07-19 & 22:19:51  & 5.6   & $G$ & $17.3\pm0.1$    & $17.9\pm0.1$ & T-CAT  &\\
ZTF21ablssud & 2021-07-19 & 23:06:29  & 5.7   & $R$ & $17.2\pm0.1$    & $17.2\pm0.3$ & Montarrenti  &\\
ZTF21ablssud & 2021-07-19 & 23:07:38  & 5.7   & $B$ & $17.8\pm0.1$    & $18.0\pm0.3$ & Montarrenti  &\\
ZTF21ablssud & 2021-07-20 & 20:15:05  & 6.5   & Clear & $17.0\pm0.1$    & $16.7\pm0.1$ & ZnithObs  &\\
ZTF21ablssud & 2021-07-20 & 20:45:31  & 6.6   & Clear & $17.3\pm0.1$    & $16.9\pm0.1$ & ZnithObs  &\\
ZTF21ablssud & 2021-07-20 & 21:08:33  & 6.6   & $R$ & $17.2\pm0.1$    & $17.3\pm0.2$ & Montarrenti  &\\
ZTF21ablssud & 2021-07-20 & 21:29:33  & 6.6   & Clear & $16.9\pm0.1$    & $16.7\pm0.1$ & T40-A77DAU  &\\
ZTF21ablssud & 2021-07-20 & 21:38:00  & 6.6   & $L$ & $17.5\pm0.1$    & $16.5\pm0.1$ & K26  &\\
ZTF21ablssud & 2021-07-20 & 22:48:24  & 6.7   & $R$ & $17.3\pm0.1$    & $17.6\pm0.2$ & Montarrenti  &\\
ZTF21ablssud & 2021-07-20 & 21:41:22  & 8.6   & Clear & $17.5\pm0.1$    & $17.1\pm0.1$ & T40-A77DAU  &\\
ZTF21ablssud & 2021-07-23 & 00:39:53  & 8.7   & $V$ & $17.7\pm0.1$    & $17.6\pm0.1$ & T-BRO  &\\
ZTF21ablssud & 2021-07-23 & 00:49:00  & 8.7   & $R$ & $17.4\pm0.1$    & $17.3\pm0.1$ & T-BRO  &\\
ZTF21ablssud & 2021-07-23 & 00:58:07  & 8.7   & $I$ & $17.2\pm0.2$    & $17.2\pm0.1$ & T-BRO  &\\
ZTF21ablssud & 2021-07-23 & 01:14:57  & 8.8   & Clear & $17.5\pm0.1$    & $17.2\pm0.1$ & T-BRO  &\\
ZTF21ablssud & 2021-07-24 & 07:16:06  & 10.0   & $r^\prime$   & -                  & $17.8\pm0.1$   & ZTF  \\
ZTF21ablssud & 2021-07-24 & 08:39:27  & 10.1   & $g^\prime$   & -                  & $18.0\pm0.1$   & ZTF  \\
ZTF21ablssud & 2021-07-27 & 21:51:12  & 13.6   & $R$ & $17.8\pm0.1$    & $17.5\pm0.3$ & Montarrenti  &\\
ZTF21ablssud & 2021-07-28 & 07:12:16  & 14.0   & $r^\prime$   & -                  & $18.1\pm0.1$   & ZTF  \\
ZTF21ablssud & 2021-07-28 & 08:05:58  & 14.0   & $g^\prime$   & -                  & $18.2\pm0.1$   & ZTF  \\
ZTF21ablssud & 2021-07-29 & 21:16:28  & 15.6   & $B$ & $18.8\pm0.1$    & $18.8\pm0.2$ & Montarrenti  &\\
ZTF21ablssud & 2021-07-29 & 21:17:37  & 15.6   & $R$ & $18.1\pm0.2$    & $18.0\pm0.1$ & Montarrenti  &\\
ZTF21ablssud & 2021-07-30 & 07:59:11  & 16.0   & $r^\prime$   & -                  & $18.2\pm0.1$   & ZTF  \\
ZTF21ablssud & 2021-07-30 & 08:40:48  & 16.1   & $g^\prime$   & -                  & $18.6\pm0.1$   & ZTF  \\
ZTF21ablssud & 2021-08-02 & 00:42:03 &   18.7  & $B$ & $19.0\pm0.1$  &    $19.0\pm0.2$ & VIRT &\\ 
ZTF21ablssud & 2021-08-02 & 04:01:46 &   18.9  & $R$ & $18.4\pm0.1$  &    $18.0\pm0.1$               & VIRT &\\ 
ZTF21ablssud & 2021-08-02 & 06:15:48  & 19.0   & $g^\prime$   & -                  & $18.9\pm0.1$   & ZTF  \\
ZTF21ablssud & 2021-08-02 & 10:32:43  & 19.1   & $R$ & $18.3\pm0.1$    & $18.2\pm0.1$ & TRT-SRO &\\ 
ZTF21ablssud & 2021-08-04 & 05:38:52  & 20.9   & $r^\prime$   & -                  & $18.7\pm0.1$   & ZTF  \\
ZTF21ablssud & 2021-08-04 & 07:12:07  & 21.0   & $g^\prime$   & -                  & $19.0\pm0.1$   & ZTF  \\
ZTF21ablssud & 2021-08-05 & 04:53:24 & 21.9& $I$ & $18.3\pm0.1$    &  $18.6\pm0.2$ & TRT-SRO &\\ 
ZTF21ablssud & 2021-08-06 & 05:14:32 & 22.9& $I$ & $18.5\pm0.1$    & $18.6\pm0.1$ & TRT-SRO &\\ 
ZTF21ablssud & 2021-08-09 & 05:59:45  & 25.9   & $r^\prime$   & -                  & $18.9\pm0.1$   & ZTF  \\
ZTF21ablssud & 2021-08-09 & 22:48:05 & 26.7& $I$ & $18.4\pm0.2$   & $18.2\pm0.1$ & UBAI/NT-60 &\\
ZTF21ablssud & 2021-08-09 & 22:54:20 &    26.7     & $R$ & $18.5\pm0.2$ &   $18.5\pm0.1$    &  UBAI/NT-60 & \\ 
ZTF21ablssud & 2021-08-10 & 21:42:37 &   27.6      & $I$ & $18.8\pm0.2$ &    $18.6\pm0.1$           &  UBAI/NT-60 & \\
\end{tabular}
\end{table*}

\begin{table*}
\addtocounter{table}{-1}
\caption{Continued.}
\begin{tabular}{cccccccccl}
Source &  Obs date & Time & $\delta \mathrm{t}$ (days) & filter & Detec. mag (STDpipe) & Muphoten (or ZTF) & Telescope/Observer & \\
\hline
ZTF21abotose &  2021-07-28 & 05:10:28 & 0.0   & $g^\prime$   & -                  & $18.7\pm0.1$   & ZTF  & \\
ZTF21abotose &  2021-07-30 & 04:41:20 & 2.0   & $r^\prime$   & -                  & $19.1\pm0.1$   & ZTF  & \\
ZTF21abotose &  2021-07-30 & 05:28:22 & 2.0   & $g^\prime$   & -                  & $19.3\pm0.1$   & ZTF  & \\
ZTF21abotose &  2021-07-30  & 20:05:42 &  2.6 & $R$  &  $18.7\pm0.1$    &  $18.7\pm0.1$  & Montarrenti & \\
ZTF21abotose &  2021-07-30  & 21:13:22 &  2.7 & Clear  &  $19.4\pm0.2$  & $18.8\pm0.1$  & T40-A77DAU & \\
ZTF21abotose &  2021-07-30 & 04:39:44 & 4.0   & $r^\prime$   & -                  & $19.7\pm0.2$   & ZTF  & \\
ZTF21abotose &  2021-07-30 & 05:36:13 & 4.0   & $g^\prime$   & -                  & $19.8\pm0.2$   & ZTF  & \\
ZTF21abotose &  2021-08-01  & 20:50:06 &  4.7 & Clear  &  $19.2\pm0.1$    &  $19.3\pm0.1$  & MSXD-A77 & \\
ZTF21abotose &  2021-08-08 &  06:06:33 & 6.1   & $g^\prime$   & -                  & $20.2\pm0.2$   & ZTF  & \\
ZTF21abotose &  2021-08-04  & 21:44:34 &  7.7 & $B$  &  $20.1\pm0.1$    &  $20.2\pm0.1$  &  T-CAT   & \\
ZTF21abotose &  2021-08-04  & 21:44:34 &  7.7 & $R$  &  $19.6\pm0.1$    &  $19.5\pm0.1$  &  T-CAT   & \\
ZTF21abotose &  2021-08-06  & 22:09:01 &  9.7 & $B$  & $19.8\pm0.1$   &  $19.9\pm0.1$  &  T-CAT   & \\
ZTF21abotose &  2021-08-06  & 22:09:01 &  9.7 & $G$  & $19.8\pm0.1$   &  $20.3\pm0.1$  &  T-CAT   & \\
ZTF21abotose &  2021-08-06  & 22:09:01 &  9.7 & $R$  & $19.2\pm0.1$   &  $19.3\pm0.1$  &  T-CAT   & \\
ZTF21abotose &  2021-08-08  & 22:49:42 &  11.8 & $B$  & $19.6\pm0.1$   &  $19.7\pm0.1$  &  T-CAT   & \\
ZTF21abotose &  2021-08-08  & 22:49:42 &  11.8 & $G$  & $19.4\pm0.1$   &  $19.9\pm0.1$  &  T-CAT   & \\
ZTF21abotose &  2021-08-08  & 22:49:42 &  11.8 & $R$  & $19.0\pm0.1$   &  $19.0\pm0.1$  &  T-CAT   & \\
ZTF21abotose &  2021-08-09  & 21:55:18 &  12.7 & $B$  & $19.4\pm0.1$   &  $19.5\pm0.1$  &  T-CAT   & \\
ZTF21abotose &  2021-08-09  & 21:55:18 &  12.7 & $G$  & $19.4\pm0.1$   &  $19.5\pm0.1$  &  T-CAT   & \\
ZTF21abotose &  2021-08-09  & 21:55:18 &  12.7 & $R$  & $18.9\pm0.1$   &  $19.0\pm0.1$  &  T-CAT   & \\
ZTF21abotose &  2021-08-10 &  05:08:49 & 13.0   & $r^\prime$   & -                  & $19.0\pm0.2$   & ZTF  & \\
ZTF21abotose &  2021-08-10 &  05:39:08 & 13.0   & $g^\prime$   & -                  & $19.2\pm0.2$   & ZTF  & \\
ZTF21abotose &  2021-08-10  & 22:05:26 &  13.7 & $B$  & $19.6\pm0.1$   &  $19.6\pm0.1$  &  T-CAT   & \\
ZTF21abotose &  2021-08-10  & 22:05:26 &  13.7 & $G$  & $19.3\pm0.1$   &  $19.4\pm0.1$  &  T-CAT   & \\
ZTF21abotose &  2021-08-10  & 22:05:26 &  13.7 & $R$  & $18.8\pm0.1$   &  $18.7\pm0.1$  &  T-CAT   & \\
ZTF21abotose &  2021-08-11  & 20:45:01 &  14.7 & $B$  & $19.4\pm0.1$   &  $19.3\pm0.1$  &  T-CAT   & \\
ZTF21abotose &  2021-08-13 &  04:14:25 & 16.0   & $g^\prime$   & -                  & $19.2\pm0.2$   & ZTF  & \\
ZTF21abotose &  2021-08-13  &  21:50:45 &  16.7 & $G$  & $18.8\pm0.1$   &  $19.1\pm0.1$  &  T-CAT   & \\
ZTF21abotose &  2021-08-13  &  21:50:45 &  16.7 & $R$  & $18.7\pm0.1$   &  $18.9\pm0.1$  &  T-CAT   & \\
ZTF21abotose &  2021-08-15 &  04:09:00 & 18.0   & $r^\prime$   & -                  & $18.6\pm0.1$   & ZTF  & \\
ZTF21abotose &  2021-08-15 &  05:52:39 & 18.0   & $g^\prime$   & -                  & $19.0\pm0.2$   & ZTF  & \\
ZTF21abotose &  2021-08-15 &  05:52:39 & 31.1   & $r^\prime$   & -                  & $19.1\pm0.2$   & ZTF  & \\
ZTF21abotose &  2021-08-29  &  21:16:13 &  32.7 & $B$  & $20.8\pm0.1$   &  $20.9\pm0.1$  &  T-CAT   & \\
ZTF21abotose &  2021-08-29  &  21:16:13 &  32.7 & $G$  & $19.7\pm0.1$   &  $20.5\pm0.1$  &  T-CAT   & \\
ZTF21abotose &  2021-08-29  &  21:16:13 &  32.7 & $R$  & $19.3\pm0.1$   &  $19.3\pm0.1$  &  T-CAT   & \\
ZTF21abotose &  2021-08-31 &  05:19:56 & 34.0   & $g^\prime$   & -                  & $20.1\pm0.3$   & ZTF  & \\
\hline
ZTF21absvlrr & 2021-08-12 & 09:52:43 & 0.0   & $g^\prime$   & -                  & $18.7\pm0.20$   & ZTF  & \\
ZTF21absvlrr & 2021-08-12 & 11:05:04 & 0.1   & $r^\prime$   & -                  & $18.6\pm0.10$   & ZTF  & \\
ZTF21absvlrr & 2021-08-13  & 09:23:51 & 1.0 & $r^\prime$ &  $18.5\pm0.1$  &   $18.5\pm0.1$   & T-PDA \\
ZTF21absvlrr & 2021-08-13  & 12:18:58 & 1.1 & Clear &  $18.0\pm0.1$  & $17.4\pm0.1$  &    Beverly-Begg \\
ZTF21absvlrr & 2021-08-14 & 02:01:14 & 1.7 & $B$  & $18.0\pm0.2$   &    $17.9\pm0.1$  & Montarrenti \\
ZTF21absvlrr & 2021-08-14 & 02:02:22 & 1.7 &  $R$ &  $17.7\pm0.1$  &    $17.5\pm0.1$   & Montarrenti \\
ZTF21absvlrr & 2021-08-14 & 10:02:31 & 2.0  & $g^\prime$   & -                  & $17.6\pm0.1$   & ZTF  & \\
ZTF21absvlrr & 2021-08-14 & 11:04:16 & 2.0  & $r^\prime$   & -                  & $17.7\pm0.1$   & ZTF  & \\
ZTF21absvlrr & 2021-08-15 & 03:13:25 & 2.7 &   Clear  & $17.3\pm0.1$  & $16.6\pm0.1$  & T40-A77DAU \\
ZTF21absvlrr & 2021-08-16 & 10:08:03 & 4.0  & $r^\prime$   & -                  & $17.0\pm0.1$   & ZTF  & \\
ZTF21absvlrr & 2021-08-17 & 12:50:20 & 4.1   & $V$ & $16.7\pm0.1$    & $16.6\pm0.1$ & TRT-SBO & \\ 
ZTF21absvlrr & 2021-08-17 & 11:03:01 & 5.1  & $r^\prime$   & -                  & $16.7\pm0.1$   & ZTF  & \\
ZTF21absvlrr & 2021-08-19 & 10:38:15 & 7.0   & $R$ & $16.1\pm0.1$    & $16.2\pm0.2$ & TRT-SBO & \\
ZTF21absvlrr & 2021-08-20 & 09:30:58 & 8.0  & $r^\prime$   & -                  & $15.7\pm0.1$   & ZTF  & \\
ZTF21absvlrr & 2021-08-22 & 08:37:29 & 10.0  & $g^\prime$   & -                  & $16.1\pm0.1$   & ZTF  & \\
ZTF21absvlrr & 2021-08-27 & 02:24:58 & 14.7 & $B$  &   $15.7\pm0.1$  &    $15.6\pm0.2$ & SUTO \\
ZTF21absvlrr & 2021-08-27 & 02:30:30 & 14.7  & $V$  &  $15.6\pm0.1$   &  $15.6\pm0.1$  &  SUTO  \\
ZTF21absvlrr & 2021-08-27 & 02:36:06 & 14.7  & $R$ &   $15.6\pm0.1$   &  $15.6\pm0.1$  &  SUTO  \\
\end{tabular}
\end{table*}

\begin{table*}
\addtocounter{table}{-1}
\caption{Continued.}
\begin{tabular}{cccccccccl}
Source &  Obs date & Time & $\delta \mathrm{t}$ (days) & filter & Detec. mag (STDpipe) & Muphoten (or ZTF) & Telescope/Observer & \\
\hline
ZTF21absvlrr & 2021-08-17 & 13:34:02 & 5.2   & $R$ & $16.5\pm0.1$    & $16.5\pm0.1$ & TRT-SBO & \\ 
ZTF21absvlrr & 2021-08-17 & 13:54:27 & 5.2   & $I$ & $16.8\pm0.1$    & $16.8\pm0.1$ & TRT-SBO & \\ 
ZTF21absvlrr & 2021-08-27 & 11:05:01 & 15.1  & $r^\prime$   & -                  & $15.5\pm0.1$   & ZTF  & \\
ZTF21absvlrr & 2021-08-28 & 02:26:24 & 15.7  & $V$  &   $15.7\pm0.1$ & $15.6\pm0.2$   &  SUTO  \\
ZTF21absvlrr & 2021-08-28 & 08:34:36 & 16.0  & $g^\prime$   & -                  & $15.3\pm0.1$   & ZTF  & \\
ZTF21absvlrr & 2021-08-28 & 10:03:56 & 16.0  & $r^\prime$   & -                  & $15.5\pm0.1$   & ZTF  & \\
ZTF21absvlrr & 2021-10-03 & 06:54:26 & 51.9  & $r^\prime$   & -                  & $17.7\pm0.1$   & ZTF  & \\
ZTF21absvlrr & 2021-10-03 & 08:03:41 & 51.9  & $g^\prime$   & -                  & $16.8\pm0.1$   & ZTF  & \\
ZTF21absvlrr & 2021-10-07 & 20:28:50 & 56.5 & $V$  & $16.8\pm0.1$ & $17.0\pm0.2$ & Terskol/Zeiss-600 &\\ 
ZTF21absvlrr & 2021-10-07 & 20:29:41 & 56.5 & $R$ & $16.5\pm0.1$ & $16.5\pm0.2$ & Terskol/Zeiss-600 &\\ 
ZTF21absvlrr & 2021-10-07 & 20:30:31 & 56.5 & $I_c$  & $16.1\pm0.1$ & $16.3\pm0.2$ & Terskol/Zeiss-600 &\\
ZTF21absvlrr & 2021-10-08 & 22:12:08 & 57.5 & $V$ & $17.3\pm0.1$   & $17.4\pm0.2$   & Terskol/Zeiss-600 &\\ 
ZTF21absvlrr & 2021-10-08 & 22:12:38 & 57.5 & $R$ & $16.9\pm0.1$  & $17.1\pm0.2$ & Terskol/Zeiss-600 &\\ 
ZTF21absvlrr & 2021-10-08 & 22:13:09 & 57.5 & $I$ & $16.5\pm0.1$ & $16.5\pm0.2$ & Terskol/Zeiss-600 &\\ 
ZTF21absvlrr & 2021-10-10 & 08:32:18 & 59.0 & $g^\prime$   & -                  & $17.9\pm0.1$   & ZTF  & \\
\hline
ZTF21abultbr & 2021-08-20 & 11:48:36 & 0.0 & $r^\prime$& - & $18.8\pm0.2$ & ZTF  & \\
ZTF21abultbr & 2021-08-21  & 02:44:39  & 0.6  & $R$ & $18.7\pm0.1$  & $18.8\pm0.1$ & T-CAT & &   \\
ZTF21abultbr & 2021-08-21 & 02:44:39 & 0.6 & $G$ & $18.6\pm0.1$ & $18.6\pm0.1$ & T-CAT & &   \\
ZTF21abultbr & 2021-08-21  &03:32:27 & 0.6  & Clear  & $18.6\pm0.2$ & $18.2\pm0.1$& T40-A77DAU & \\
ZTF21abultbr & 2021-08-23 & 11:45:21 & 3.0 & $r^\prime$& - & $18.7\pm0.2$ & ZTF  &   \\
\hline
ZTF21acceboj & 2021-09-14 & 11:04:25 & 0.0 & $r^\prime$ & - & $18.4\pm0.1$ &  ZTF \\
ZTF21acceboj & 2021-09-15 & 22:12:32 & 1.4  & $g^\prime$ & $17.9\pm0.1$ & $17.9\pm0.1$   &  BJP/ALi-50 \\
ZTF21acceboj & 2021-09-15 & 22:20:25 & 1.4  & $r^\prime$ & $18.0\pm0.1$ & $17.8\pm0.1$   &  BJP/ALi-50 \\ 
ZTF21acceboj & 2021-09-16 & 11:02:48 & 2.0 & $r^\prime$ & - & $17.8\pm0.1$ &  ZTF \\
ZTF21acceboj & 2021-09-16 & 11:33:36 & 2.0 & $g^\prime$ & - & $17.8\pm0.1$ &  ZTF \\
ZTF21acceboj & 2021-10-02 & 11:01:57 & 18.0 & $r^\prime$ & - & $17.4\pm0.1$ &  ZTF \\
ZTF21acceboj & 2021-10-02 & 11:33:58 & 18.0 & $g^\prime$ & - & $17.8\pm0.1$ &  ZTF \\
ZTF21acceboj & 2021-10-07 & 23:48:35 & 23.5 & $V_c$ & $18.0\pm0.1$ & $18.3\pm0.3$ & Terskol/Zeiss-600 \\ 
ZTF21acceboj & 2021-10-07 & 23:49:26 & 23.5  & $R_c$ & $17.2\pm0.2$ & $17.2\pm0.2$ & Terskol/Zeiss-600 \\
ZTF21acceboj & 2021-10-07 & 23:50:17 & 23.5 & $I_c$ & $16.5\pm0.1$ & $16.6\pm0.3$ & Terskol/Zeiss-600 \\ 
ZTF21acceboj & 2021-10-09 & 00:23:44 & 24.5 & $V$ &  $18.5\pm0.1$ & $18.5\pm0.1$ & Terskol/Zeiss-600 \\ 
ZTF21acceboj & 2021-10-09 & 00:24:15 & 24.5 & $R_c$ & $17.5\pm0.1$ & $17.7\pm0.1$ & Terskol/Zeiss-600 \\
ZTF21acceboj & 2021-10-09 & 00:24:46 & 24.5 & $I_c$ & $17.2\pm0.2$ & $17.3\pm0.2$ & Terskol/Zeiss-600 \\ 
ZTF21acceboj & 2021-10-09 & 09:30:38 & 24.9 & $g^\prime$ & - & $18.0\pm0.1$ &  ZTF \\
ZTF21acceboj & 2021-10-10 & 00:49:13 & 25.5 & $V$ & $17.9\pm0.1$  & $18.0\pm0.1$ & Terskol/Zeiss-600 \\ 
ZTF21acceboj & 2021-10-10 & 00:49:53 & 25.5 & $R_c$ & $17.4\pm0.1$ & $17.5\pm0.1$ & Terskol/Zeiss-600 \\ 
ZTF21acceboj & 2021-10-10 & 00:50:34 & 25.5 & $I_c$ & $17.1\pm0.1$ & $16.9\pm0.2$ & Terskol/Zeiss-600 \\ 
ZTF21acceboj & 2021-10-11 & 10:31:51 & 26.9 & $g^\prime$ & - & $18.0\pm0.1$ &  ZTF \\
ZTF21acceboj & 2021-10-11 & 11:34:57 & 27.0 & $r^\prime$ & - & $17.5\pm0.1$ &  ZTF \\
 \hline
\end{tabular}
\end{table*}

\section{Linear fitting regression} \label{appendix_linearfit}

The maximum likelihood approach is used to constrain the nature of the  rapidly evolving transient phenomena.
Maximum likelihood estimation (MLE) is a general method for estimating the best-fitting  parameters $(\theta)$ of a statistical model, under the assumption that the true distribution of these parameters  is known. Fitting any mathematical form of the model, these estimates maximize the probability of the observed values across a series of observations of a random variable. We use the concept of a sufficient statistic ($\text{S}_{tt}$, $\text{S}_{y})$ which contains all the information from the data that is needed to find the best-fitting $\theta$ parameters.
Our maximum likelihood estimation function is defined as follows:

\begin{align*}
L(\theta) &= L(a,b)\\
          &= \log P(\text{y}|a,b,H) \\
          & = 
 -\frac{1}{2}\sum_i \log{2\pi\sigma_i^2} - \frac{1}{2} \sum_i \frac{(\text{y}_i - a\text{t}_i - b)^2}{\sigma_i^2}
\end{align*}
Where, $\sigma_i$ is the error of the measure of the magnitude $\text{y}_i$ at the date $\text{t}_i$. 

The sufficient statistics, $\text{S}_{tt} $ and $\text{S}_{y} $, are zeros of the partial first derivatives of log maximum likelihood $L(\theta)$.

$$ \frac{\partial \log L}{\partial \theta} = 0 $$
This gives us :
$$
\left\{
    \begin{array}{ll}

       \text{S}_{tt} = a\sum_i \frac{\text{t}_i^{2}}{\sigma_i^2} + b\sum_i \frac{\text{t}_i}{\sigma_i^2}  = \sum_i \frac{t_i y_i}{\sigma_i^2} 
 
     \\
     \\
     
  \text{S}_{y} = a\sum_i \frac{\text{t}_i}{\sigma_i^2} + b\sum_i \frac{1}{\sigma_i^2}  = \sum_i \frac{y_i}{\sigma_i^2} 
     \end{array}
\right.
$$

The best-fitting estimators ($a, b$) are given following the resolution of  these linear equations above. And our linear regression is defined by $\text{Y}_i = a\text{t}_i + b$. The parameter $a$ is our light linear evolution  rate. We define a \textbf{slow transient}, a transient whose decay rate is either negative (the brightness rises) or below 0.3 mag/day in a given filter.


\section{Affiliations}
$^{1}$E. Kharadze Georgian National Astrophysical Observatory, Mt.Kanobili, Abastumani, 0301, Adigeni, Georgia; Ilia State University, Kakutsa Cholokashvili ave 3/5, Tbilisi 0162, Georgia\\
$^{2}$Samtskhe-Javakheti  State  University, Rustaveli Str. 113,  Akhaltsikhe, 0080,  Georgia\\
$^{3}$American University of Sharjah, Physics Department, PO Box 26666, Sharjah, UAE\\
$^{4}$Artemis, Observatoire de la Côte d’Azur, Université Côte d’Azur, Boulevard de l'Observatoire, 06304 Nice, France\\
$^{5}$GRAPPA, Anton Pannekoek Institute for Astronomy and Institute of High-Energy Physics, University of Amsterdam, Science Park 904,1098 XH Amsterdam, The Netherlands\\
$^{6}$Universit\'e de Paris, CNRS, Astroparticule et Cosmologie, F-75013 Paris, France\\
$^{7}$Astronomical Observatory\ Taras Shevshenko National University of Kyiv, Observatorna str. 3, Kyiv, 04053, Ukraine\\
$^{8}$ University of Iceland, Sæmundargata 2, 102 Reykjavík, Iceland \\
$^{9}$ Aix Marseille Univ, CNRS, CNES, LAM, IPhU, Marseille, France\\
$^{10}$Société Astronomique de Lyon, 9 Avenue Charles Andr\'e, Saint Genis Laval, France\\ 
$^{11}$ICAMER Observatory of NAS of Ukraine 27 Acad. Zabolotnoho Str., Kyiv, 03143, Ukraine\\
$^{12}$Instituto de Astrof\'isica de Andaluc\'ia (IAA-CSIC), Glorieta de la Astronom\'ia s/n, 18008 Granada, Spain\\
$^{13}$Université Grenoble-Alpes, Universit\'e Savoie Mont Blanc, CNRS/IN2P3 Laboratoire d'Annecy-le-Vieux de Physique des particules, France\\
$^{14}$ Observatoire de Dauban, 04 Banon - France\\
$^{15}$Groupe Astronomique de Querqueville, 61, rue Roger Glinel, 50460, Cherbourg en cotentin, France\\
$^{16}$Vereniging Voor Sterrenkunde,  Balen-Neetlaan 18A, B-2400, Mol, Belgium \\
$^{17}$Zeeweg 96, B-8200 Brugge, Belgium \\
$^{18}$Ulugh Beg Astronomical Institute, Uzbekistan Academy of Sciences, Astronomy str. 33, Tashkent 100052, Uzbekistan\\
$^{19}$MP-G2A, Midi Pyr\'en\'ees, Groud Astrophysic Addict, 81570, Cuq, France\\
$^{20}$ El Center for Backyard Astrophysics, Spain\\
$^{21}$School of Physics and Astronomy, University of Minnesota, Minneapolis, Minnesota 55455, USA\\
$^{22}$Institute for Physics and Astronomy, University of Potsdam, D-14476 Potsdam, Germany\\
$^{23}$Max Planck Institute for Gravitational Physics (Albert Einstein Institute), Am M{\"u}hlenberg 1, D-14476\\
$^{24}$OHP, Observatoire de Haute-Provence, CNRS, Aix Marseille University, Institut Pythéas, St Michel l’Observatoire, France\\
$^{25}$IJCLab, Univ Paris-Saclay, CNRS/IN2P3, Orsay, France\\
$^{26}$Institut d’Astrophysique de Paris, 98 bis boulevard Arago, 75014 Paris France\\
$^{27}$Institute of Earth Systems, University of Malta, MSD 2080, Malta \\  
$^{28}$Znith Observatory, Naxxar, Malta \\  
$^{29}$APPAM, Montredon-Labessonnié, France\\
$^{30}$University of the Virgin Islands, United States Virgin Islands 00802, USA\\
$^{31}$Volkssternwarte Paderborn, Im Schloßpark 13,33104 Paderborn, Germany\\
$^{32}$Hidden Valley Observatory, E9891 810th Ave., Colfax, WI., USA\\
$^{33}$LPC, Université Clermont Auvergne, CNES/IN2P3, F-63000, France\\
$^{34}$Dunedin Astronomical Society (DAS), Royal Astronomical Society of New Zealand\\
$^{35}$FZU - Institute of Physics of the Czech Academy of Sciences, Na Slovance 1999/2, CZ-182 21, Praha, Czech Republic\\
$^{36}$Laboratoire de Physique et de Chimie de l’Environnement, Université Joseph KI-ZERBO, Ouagadougou, Burkina Faso\\
$^{37}$IRAP, Universit\'e de Toulouse, CNRS, UPS, 14 Avenue Edouard Belin, F-31400 Toulouse, France\\
$^{38}$Universit\'e Paul Sabatier Toulouse III, Universit\'e de Toulouse, 118 route de Narbonne, 31400 Toulouse, France\\
$^{39}$ Contern Observatory, L-5316 Contern, Luxembourg \\
$^{40}$Beijing Planetarium, Beijing Academy of Science and Technology, Beijing, 100044, China \\
$^{41}$Montarrenti Observatory, S.S. 73 Ponente, I-53018, Sovicille, Siena, Italy \\
$^{42}$OPERA Z97-, 33820 Saint Palais , France\\ 
$^{43}$Observatory Uranoscope de l’Ile de France , Allee Camille Flammarion 77220 Gretz –Armainvilliers, France\\ 
$^{44}$Observatoire du "Crous des Gats", 31550 Cintegabelle, France\\
$^{45}$Centre for Astrophysics and Supercomputing, Swinburne University of Technology, Mail Number H29, PO Box 218, 31122 Hawthorn, VIC, Australia\\
$^{46}$OAR Telescope/NSF's NOIRLab, Avda Juan Cisternas 1500, 1700000, La Serena, Chile\\
$^{47}$National Astronomical Research Institute of Thailand (Public Organization), 260, Moo 4, T. Donkaew, A. Mae Rim, Chiang Mai, 50180, Thailand\\
$^{48}$OrangeWave Innovative Science, LLC, Moncks Corner, SC 29461, USA\\
$^{49}$Department of Electronics, Electrical Engineering and Microelectronics, Silesian University of Technology, Gliwice, Poland\\
$^{50}$Université de Strasbourg, CNRS, IPHC UMR 7178, F-67000 Strasbourg, France\
$^{51}$School of Physics and Astronomy, Rochester Institute of Technology, 84 Lomb Memorial Drive, Rochester, NY 14623, USA \\
$^{52}$Main Astronomical Observatory of National Academy of Sciences of Ukraine, 27 Acad. Zabolotnoho Str., Kyiv, 03143, Ukraine\\
$^{53}$Société Astronomique Populaire du Centre ,40 grande rue, 18340 Arçay, France\\
$^{54}$Universit\'e Paris-Saclay, CNRS, CEA, D\'epartement d'Astrophysique, Astrophysique, Instrumentation et Mod\'elisation de Paris-Saclay, 91191, Gif-sur-Yvette, France.\\
$^{55}$Astronomy and Space Physics Department, Taras Shevchenko National University of Kyiv, Glushkova ave., 4, Kyiv, 03022, Ukraine\\
$^{56}$National Center «Junior academy of sciences of Ukraine», 38-44, Dehtiarivska St., Kyiv, 04119, Ukraine\\
$^{57}$Institute of Gravitational Research, University of Glasgow, Glasgow, G12 8QQ \\
$^{58}$ICAstronomy, Oria, Almería, Spain \\ 
$^{59}$Deep Sky Chile, Pichasca, Rio Hurtado, Chile \\ 
$^{60}$National University of Uzbekistan, 4 University str., Tashkent 100174, Uzbekistan\\
$^{61}$INFN, Laboratori Nazionali del Sud, I-95125 Catania, Italy\\
$^{62}$Physics Department and Astronomy Department, Tsinghua University, https://fr.overleaf.com/project/60fc6455b426b75b631daf74Beijing, 100084, China\\





\label{lastpage}

\end{document}